\documentclass{article}
\pdfoutput=1
\usepackage{arxiv}

\usepackage[utf8]{inputenc} 
\usepackage[T1]{fontenc}   
\usepackage{hyperref}       
\usepackage{url}            
\usepackage{booktabs}      
\usepackage{amsfonts}       
\usepackage{nicefrac}      
\usepackage{microtype}     

\usepackage{booktabs} 
\usepackage{algorithm}
\usepackage[noend]{algorithmic}
\usepackage{enumitem}
\usepackage{subcaption}   
\usepackage{multirow}
\usepackage{graphicx}
\usepackage{amsmath,amsthm}
\usepackage{amssymb}
\usepackage{url}

\newtheorem{example}{Example}
\newtheorem{problem}{Problem}
\newtheorem{lemma}{Lemma}

\usepackage{setspace}
\title{Scalable Distributed Subtrajectory Clustering}

\author{
  Panagiotis Tampakis$^1$, Nikos Pelekis$^2$, Christos Doulkeridis$^3$ and Yannis Theodoridis$^1$\\
  $^1$Department of Informatics\\
  $^2$Department of Statistics \&\ Insurance Science\\
  $^3$Department of of Digital Systems\\
  University of Piraeus\\
  Piraeus, Greece \\
  \texttt{\{ptampak,npelekis,cdoulk,ytheod\}@unipi.gr} \\
}

\begin{document}

\newcommand{\SWITCH}[1]{\STATE \textbf{switch} (#1)}
\newcommand{\ENDSWITCH}{\STATE \textbf{end switch}}
\newcommand{\CASE}[1]{\STATE \textbf{case} #1\textbf{:} \begin{ALC@g}}
\newcommand{\ENDCASE}{\end{ALC@g}}
\newcommand{\CASELINE}[1]{\STATE \textbf{case} #1\textbf{:} }
\newcommand{\DEFAULT}{\STATE \textbf{default:} \begin{ALC@g}}
\newcommand{\ENDDEFAULT}{\end{ALC@g}}
\newcommand{\DEFAULTLINE}[1]{\STATE \textbf{default:} }
\newcommand{\prob}{{Distributed Subtrajectory Clustering}}

\maketitle

\begin{abstract}
Trajectory clustering is an important operation of knowledge discovery from mobility data. Especially nowadays, the need for performing advanced analytic operations over massively produced data, such as mobility traces, in efficient and scalable ways is imperative. However, discovering clusters of complete trajectories can overlook significant patterns that exist only for a small portion of their lifespan. In this paper, we address the problem of \emph{\prob} in an efficient and highly scalable way. The problem is challenging because the subtrajectories to be clustered are not known in advance, but they need to be discovered dynamically based on adjacent subtrajectories in space and time. Towards this objective, we split the original problem to three sub-problems, namely \emph{Subtrajectory Join}, \emph{Trajectory Segmentation} and \emph{Clustering and Outlier Detection}, and deal with each one in a distributed fashion by utilizing the MapReduce programming model. The efficiency and the effectiveness of our solution is demonstrated experimentally over a synthetic and two large real datasets from the maritime and urban domains and through comparison with two state of the art subtrajectory clustering algorithms.
\end{abstract}


\maketitle

\section{Introduction} \label{sec_intro}

Nowadays, the unprecedented rate of trajectory data generation, due to the proliferation of GPS-enabled devices, poses new challenges in terms of storing, querying, analyzing and extracting knowledge from big mobility data. 
One of these challenges is cluster analysis, which aims at identifying clusters of moving objects (thus, unveil hidden patterns of collective behavior), as well as detecting moving objects that demonstrate abnormal behaviour and can be considered as outliers. 

The research so far has focused mainly in methods that aim to identify  specific collective behavior patterns among moving objects, such as~\cite{DBLP_journals/gis/LaubeIW05, DBLP_conf/ssd/KalnisMB05, DBLP_journals/pvldb/JeungYZJS08, DBLP_journals/pvldb/Orakzai19, DBLP_journals/pvldb/LiDHK10, DBLP_journals/dke/LiBK15, DBLP_conf/icde/TangZYHLHP12, DBLP_conf/icde/ZhengZYS13, DBLP_journals/pvldb/FanZWT16}. However, this kind of approaches operate at specific predefined temporal ``snapshots'' of the dataset, thus ignoring the route of each moving object between these sampled points. Another line of research, tries to identify patterns that are valid for the entire lifespan of the moving objects~\cite{DBLP_journals/jiis/NanniP06, DBLP_journals/kais/PelekisKKFT11, DBLP_journals/cluster/DengHZHD15, DBLP_journals/ijghpc/SekiJU13}. However, discovering clusters of complete trajectories can overlook significant patterns that might exist only for some portions of their lifespan. The following motivating example shows the merits of subtrajectory clustering. 

\begin{figure}[thb]
    \centering
    \begin{subfigure}{0.4\columnwidth}
        \centering
        \includegraphics[width=\columnwidth]{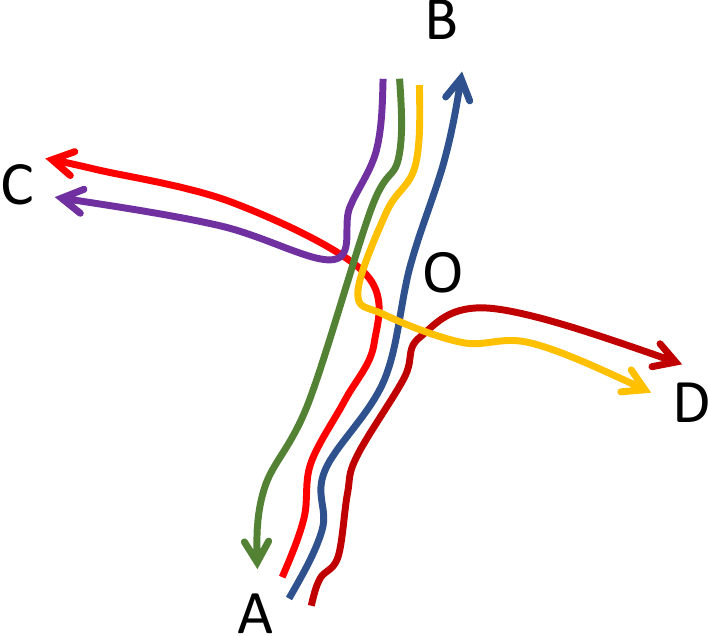}
        \caption{}
    \end{subfigure}%
    ~ 
    \begin{subfigure}{0.4\columnwidth}
        \centering
        \includegraphics[width=\columnwidth]{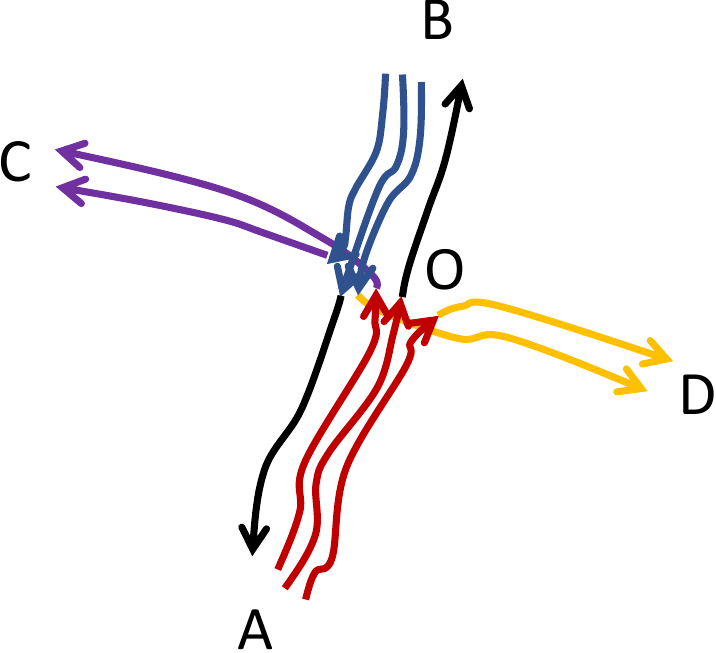}
        \caption{}
    \end{subfigure}
    \caption{(a) Six trajectories moving in the xy-plane and (b) 4 clusters (red, blue, orange and purple) and 2 outliers (black).}
\label{fig_intro}
\end{figure}

\begin{example} \label{ex:intro} \emph{(Subtrajectory clustering)}
Figure~\ref{fig_intro}(a) illustrates six trajectories moving in the xy-plane, where each one of them has a different origin-destination pair. More specifically, these pairs are $A\rightarrow{B}$, $A\rightarrow{C}$, $A\rightarrow{D}$, $B\rightarrow{A}$, $B\rightarrow{C}$ and $B\rightarrow{D}$. These six trajectories have the same starting time and similar speed. A typical trajectory clustering technique would fail to identify any clusters. However, the goal of a subtrajectory clustering method is to identify 4 clusters ($A\rightarrow{O}$ (red), $B\rightarrow{O}$ (blue), $O\rightarrow{C}$ (purple), $O\rightarrow{D}$ (orange)) and 2 outliers ($O\rightarrow{A}$ and $O\rightarrow{B}$ (black)), as depicted in Figures~\ref{fig_intro}(b).

\end{example}

The problem of subtrajectory clustering is shown to be NP-Hard (cf.~\cite{DBLP_conf/pods/AgarwalFMNPT18}). In addition, the objects to be clustered are not known beforehand (as in entire-trajectory -- from now on -- clustering algorithms), but have to be identified through a trajectory segmentation procedure. 
Efforts that try to deal with this problem in a centralized way do exist. More specifically, an approach that segments the trajectories based on their geometric features, and then clusters them by ignoring the temporal dimension is presented in~\cite{DBLP_conf/sigmod/LeeHW07}. 
Instead, the authors in~\cite{DBLP_conf/edbt/PelekisTVPT17} take into account the temporal dimension, 
and the segmentation of a trajectory takes place whenever the density of its spatiotemporal ``neighborhood'' changes significantly. 
The segmentation phase is followed by a sampling phase, where the most representative subtrajectories are selected and finally the clusters are built ``around'' these representatives. 
A similar approach is adopted in~\cite{DBLP_conf/pods/AgarwalFMNPT18}, where the goal is to identify common portions between trajectories,with respect to some constraints and/or objectives, thus taking into account the ``neighborhood'' of each trajectory. These common subtrajectories are then clustered and each cluster is represented by a pathlet, which is a point sequence that is not necessarily a subsequence of an actual trajectory.

Unfortunately, applying centralized algorithms for subtrajectory clustering (e.g.,~\cite{DBLP_conf/edbt/PelekisTVPT17, DBLP_conf/sigmod/LeeHW07, DBLP_conf/pods/AgarwalFMNPT18}) over massive data in a scalable way is far from straightforward. This calls for parallel and distributed algorithms that address the scalability requirements.
In this context, one challenge is how to partition the data in such a way so that each node can perform its computation independently, thus minimizing the communication cost between nodes, which is a cost that can turn out to be a serious bottleneck. 
Another challenge, related to partitioning, is how to achieve load balancing, in order to balance the load fairly between the different nodes.
Yet another challenge is to minimize the iterations of data processing, which are typically required in clustering algorithms.
Interestingly, there have been some recent efforts towards mining mobility data in a distributed way, such as mining co-movement patterns~\cite{DBLP_journals/pvldb/FanZWT16}, identifying frequent patterns~\cite{DBLP_journals/ijghpc/SekiJU13} or adapting already existing distributed solutions to trajectory data~\cite{DBLP_journals/cluster/DengHZHD15}, yet no approach for distributed subtrajectory clustering exists as of now.  

Motivated by these limitations, we study the \emph{\prob~(\emph{DSC})} problem, which has not been addressed yet in a scalable and efficient way.
Moreover, salient features of our approach include: (a) the discovery of clusters of subtrajectories, instead of whole trajectories, (b) spatio-temporal clustering, instead of spatial only, 
and (c) support of trajectories with variable sampling rate, length and with temporal displacement.

Our main contributions are the following: 
\begin{itemize}
 \item We formally define the problem of \emph{\prob}, investigate its properties and discuss the main challenges.
 \item We propose two neighborhood-aware trajectory segmentation algorithms, which are tailored to DSC problem, covering different application requirements.
 \item We design an efficient and scalable solution for the problem of \emph{\prob}.
 \item We perform an extensive experimental study, where the performance and the effectiveness of the proposed algorithms is evaluated by using a synthetic and two large, real trajectory datasets from different domains (urban and maritime). The merits of our solution are demonstrated with respect to two state of the art subtrajectory clustering algorithms,~\cite{DBLP_conf/edbt/PelekisTVPT17} and~\cite{DBLP_conf/sigmod/LeeHW07}.
\end{itemize}

The rest of the paper is organized as follows. In Section~\ref{sec_relat} we provide an overview of the relevant literature. Subsequently, in Section~\ref{sec_probl} we introduce the \emph{DSC} problem, in Section~\ref{sec_sol} we present our proposed solution and in Section~\ref{sec_complx} we perform a complexity analysis of the algorithms that constitute our solution. Then, in Section~\ref{sec_exper}, we present the results of our experimental study. We conclude the paper in Section~\ref{sec_concl}.

\section{Related Work} \label{sec_relat}
In recent years, an increased research interest has been observed in knowledge discovery out of mobility data. Towards this direction, several mining methods have been proposed, which can be categorized to \emph{co-movement pattern discovery}, \emph{trajectory clustering}, \emph{sequential patterns} and \emph{periodic patterns}. In this section, we focus in the first two categories of patterns that are directly related to our work.

\textbf{Co-movement patterns.}
One of the first approaches for identifying such collective mobility behavior is the so-called flock pattern~\cite{DBLP_journals/gis/LaubeIW05, DBLP_conf/gis/VieiraBT09}. 
Inspired by this, a less ``strict'' definition of flocks was proposed  in~\cite{DBLP_conf/ssd/KalnisMB05} where the notion of a moving cluster was introduced. 
There are several related works that emerged from the above ideas, like the approaches of convoys~\cite{DBLP_journals/pvldb/JeungYZJS08, DBLP_journals/pvldb/Orakzai19}, swarms~\cite{DBLP_journals/pvldb/LiDHK10}, platoons~\cite{DBLP_journals/dke/LiBK15}, traveling companion~\cite{DBLP_conf/icde/TangZYHLHP12} and gathering pattern~\cite{DBLP_conf/icde/ZhengZYS13}.

However, all of the aforementioned approaches are centralized and cannot scale to massive datasets. In this direction, the problem of efficient convoy discovery was studied both in centralized~\cite{DBLP_journals/pvldb/Orakzai19}  and distributed environment by employing the MapReduce programming model~\cite{DBLP_conf/mdm/OrakzaiCP16}. An approach that defines a new generalized mobility pattern is presented in~\cite{DBLP_journals/pvldb/FanZWT16}. In more detail, the general co-movement pattern (GCMP), is proposed, which models various co-movement patterns in a unified way and is deployed on a modern distributed platform (i.e., Apache Spark) to tackle the scalability issue. 
Even though all of these approaches provide explicit definitions of several mined patterns, their main limitation is that they search for specific collective behaviors, defined by respective parameters. Nevertheless, none of the above techniques tackles the subtrajectory clustering problem. 

\textbf{Trajectory clustering.}
Most of the aforementioned approaches operate at specific predefined temporal ``snapshots'' of the dataset, thus ignoring the route of each moving object between these ``snapshots''. Another line of research, tries to discover groups of either entire or portions of trajectories considering their routes. A typical strategy in dealing with trajectory clustering is to transform trajectories to a multi-dimensional space and then apply well-known clustering algorithms such as OPTICS~\cite{DBLP_conf/sigmod/AnkerstBKS99} and DBSCAN~\cite{DBLP_conf/kdd/EsterKSX96}. Alternatively, another approach is to define an appropriate similarity function and embed it to an extensible clustering algorithm. In this direction, there are several approaches whose goal is to group whole trajectories, including T-OPTICS~\cite{DBLP_journals/jiis/NanniP06}, that incorporates a trajectory similarity function into the OPTICS~\cite{DBLP_conf/sigmod/AnkerstBKS99} algorithm. CenTR-I-FCM~\cite{DBLP_journals/kais/PelekisKKFT11}, a variant of Fuzzy C-means, proposes a specialized similarity function that aims to tackle the inherent uncertainty of trajectory data. 

Nevertheless, trajectory clustering is a computationally intensive operation and centralized solutions cannot scale to massive datasets. In this context,~\cite{DBLP_journals/cluster/DengHZHD15} introduces a scalable GPU-based trajectory clustering approach which is based on OPTICS~\cite{DBLP_conf/sigmod/AnkerstBKS99}. 
Moreover,~\cite{DBLP_journals/ijghpc/SekiJU13} attempts to identify frequent movement patterns from the trajectories of moving objects. More specifically, they propose a MapReduce approach by employing quadtree-based hierarchical grid in order to discover complex patterns of different granularity.

\textbf{Subtrajectory clustering.} 
Nonetheless, discovering clusters of complete trajectories can overlook significant patterns that might exist only for portions of their lifespan. To deal with this, another line of research has emerged, that of \emph{Subtrajectory Clustering}. The predominant approach here is TraClus~\cite{DBLP_conf/sigmod/LeeHW07}, a partition-and-group framework for clustering 2D moving objects (i.e. TraClus ignores the time dimension) that enables the discovery of common subtrajectories. The algorithm first partitions trajectories to directed segments (i.e., subtrajectories) whenever the shape of a trajectory changes significantly, by employing the minimum description length (MDL) principle. Subsequently, the resulting subtrajectories are clustered by employing a modified version of the DBSCAN algorithm, which is applicable to directed segments. Finally, for each identified cluster the algorithm calculates a ``fictional'' representative trajectory that best describes the corresponding cluster.

A more recent approach to the problem of subtrajectory clustering, is S$^{2}$T-Clustering \cite{DBLP_conf/edbt/PelekisTVPT17}, where the goal is to partition trajectories into subtrajectories and then form groups of similar ones, while, at the same time, separate the ones that fit into no group, called outliers. It consists of two phases: a Neighborhood-aware Trajectory Segmentation (\emph{NaTS}) phase and a Sampling, Clustering and Outlier (\emph{SaCO}) detection phase. In \emph{NaTS} the trajectories are split to subtrajectories by applying a voting and segmentation process that detects homogenized subtrajectories w.r.t. the density of their neighborhood, in contrast with TraClus where the splitting is performed based on the geometric attributes of the trajectories. In \emph{SaCO} the most representative subtrajectories are selected to serve as the seeds of the clusters, around which the clusters are formed (also, the outliers are isolated). 
A slightly different approach is presented in QuT-Clustering~\cite{DBLP_journals/datamine/PelekisTVDT17} and~\cite{DBLP_conf/icde/TampakisPAAFT18}, where the goal is, given a a temporal period of interest $W$, to efficiently retrieve the clusters and outliers at subtrajectory level, that temporally intersect $W$. In order to achieve this, a hierarchical structure, called ReTraTree (for Representative Trajectory Tree) that effectively indexes a dataset for subtrajectory clustering purposes, is built and utilized.

An alternative viewpoint to the problem of subtrajectory clustering is presented in~\cite{DBLP_conf/pods/AgarwalFMNPT18}, where the goal is to identify ``common'' portions between trajectories, w.r.t. some constraints and/or objectives, cluster these ``common'' subtrajectories and represent each cluster as a pathlet, which is a point sequence that is not necessarily a subsequence of an actual trajectory. A pathlet can be viewed as a portion of a path that is traversed by many trajectories. In order to solve this problem, the authors in~\cite{DBLP_conf/pods/AgarwalFMNPT18} prove that this problem is NP-Hard and propose some approximation algorithms with theoretical guarantees, concerning the quality of the solution and the running time. Similarly, in~\cite{DBLP_conf/mdm/ZygourasG18} the goal is to identify corridors, which are frequent routes traversed by a significant number of moving objects. As already mentioned, all of the above subtrajectory clustering approaches are centralized and cannot scale to the size of today's trajectory data.

\section{Problem Formulation} \label{sec_probl}

Given a set $D$ of moving object trajectories, a trajectory $r \in D$ is a sequence of timestamped locations $\{r_1,\dots,r_N\}$. Each $r_i = (loc_i, t_i)$ represents the $i$-th sampled point, $i \in {1,\dots,N}$ of trajectory $r$, where $N$ denotes the length of $r$ (i.e. the number of points it consists of). Moreover, $loc_i$ denotes the spatial location (2D or 3D) and $t_i$ the time coordinate of point $r_i$, respectively. A subtrajectory $r_{i,j}$ is a sub-sequence $\{r_i,\dots,r_j\}$ of $r$ which represents the movement of the object between $t_i$ and $t_j$ where $i<j$ and $i,j \in {1,\dots,N}$. Let $d_{s}(r_i, s_j)$ denote the spatial distance between two points $r_i \in r$, $s_j \in s$. In our case we adopted the Euclidean distance, however, other metric distance functions might be applied. Also, let $d_{t}(r_i, s_j)$ denote the temporal distance, defined as $|r_i.t - s_j.t|$. Furthermore, let $\Delta t_{r}$ symbolize the duration of trajectory $r$ (similarly for subtrajectories).

\subsection{Similarity between (sub)trajectories} \label{sec_traj_sim}

Subtrajectory clustering relies on the use of a similarity function between subtrajectories. Although various similarity measures have been defined in literature, our choice of similarity function is motivated by the following (desired) requirements:

\begin{itemize}[topsep=2pt,leftmargin= .1in]
    \item[]\textbf{Variable sampling rate and lack of alignment.} We make the realistic assumption that the trajectories do not have a fixed sampling rate and that different trajectories might not report their position at the same timestamp.
    \item[]\textbf{Variable trajectory length.} We also assume that different trajectories might have different length (i.e. number of samples). This specification excludes euclidean-based similarity measures which deal with trajectories of equal length.
    \item[]\textbf{Temporal displacement.} A property that a desired similarity measure for (sub)trajectory clustering should hold, is to allow trajectories that have some temporal displacement to participate to the same cluster.
    \item[]\textbf{Symmetry.}  Given a pair of (sub)trajectories $r$ and $s$, an appropriate similarity measure between $r$ and $s$ should have the property of symmetry (i.e. $Sim(r,s)$=$Sim(s,r)$).
    \item[]\textbf{Efficiency.} The computation of the similarity should be efficient enough in order to be able to deal with massive volumes of data, without compromising the quality of the results.
\end{itemize}

In order to meet with the aforementioned specifications we utilize the Longest Common Subsequence (LCSS) for trajectories, as defined in~\cite{DBLP_conf/icde/VlachosGK02}. However, other trajectory similarity functions, which meet with the specifications set, are also applicable. More specifically, the LCSS utilizes two parameters, the parameter $\epsilon_t$ indicating the temporal range wherein the method searches to match a specific point, and the $\epsilon_{sp}$ parameter which is a distance threshold to indicate whether two points match or not. Hence, the similarity between two (sub)trajectories $r$ and $s$ is defined as:

\begin{equation} \label{eqn:lcss}
Sim(r, s) = \frac{LCSS_{\epsilon_{t},\epsilon_{sp}}(r,s)}{min(|r|,|s|)}
\end{equation}

\noindent where $min(|r|,|s|)$ is the length of the longest common subsequence. Moreover, it holds that $Sim(r,s) = Sim(s,r)$. 

However, LCSS returns the length of the longest common subsequence, which means that for a given point $r_i \in r$ that is matched with a specific point $s_j \in s$ the LCSS will consider the similarity between $r_i$ and $s_j$ as 1, regardless of their actual distance $d_{s}(r_i, s_j)$, which could vary from 0 to $\epsilon_{sp}$.
Put differently, LCSS considers as equally similar all the points that exist within an $\epsilon_{sp}$ range from $r$, which is a fact that might compromise the quality of the clustering results. Ideally, given two matching points $r_i \in r$ and $s_j \in s$,  $s_j$ ($r_i$, respectively) should contribute to $LCSS_{\epsilon_{t},\epsilon_{sp}}(r,s)$, proportionally to the distance $d_{s}(r_i, s_j)$. For this reason, we propose a ``weighted'' LCSS similarity between trajectories, that incorporates the aforementioned distance proportionality. In more detail, for each discovered longest common subsequence the similarity is defined as:

\begin{equation} \label{eqn:sim_traj}
Sim(r, s) = \frac{\sum\limits_{k = 1}^{min(|r|,|s|)}(1 - {\frac{d_{s}(r_k, s_k)}{\epsilon_{sp}}})}{min(|r|,|s|)}
\end{equation}

\noindent where $(r_{k},s_{k})$ is a pair of matched points.

\subsection{A Closer Look to the Subtrajectory Clustering Problem} \label{sec_st-clust_prop}

Our approach to subtrajectory clustering splits the problem in three steps.
The first step is to retrieve for each trajectory $r \in D$, all the moving objects, with their respective portion of movement, that moved close enough in space and time with $r$, for at least some time duration. Actually, this first step is a well-defined problem in the literature of mobility data management, known as \emph{subtrajectory join}, and more specifically the case of self-join. In detail, the subtrajectory join will return for each pair of (sub)trajectories, all the common subsequencies that have at least some time duration, which are actually candidates for the longest common subsequence. Formally:

\begin{problem} \label{prb_trj_join}
\textbf{(Subtrajectory Join)}
Given a temporal tolerance $\epsilon_t$, a spatial threshold $\epsilon_{sp}$ and a time duration $\delta t$, retrieve all pairs of subtrajectories $(r',s') \in D$ such that: (a) for each pair $\Delta t_{r'},\Delta t_{s'} \geq \delta t$, (b) $\forall r_i \in r'$ there exists at least one $s_j \in s'$ so that $d_{s}(r_i, s_j) \leq \epsilon_{sp}$ and $d_{t}(r_i, s_j) \leq \epsilon_t$, and (c) $\forall s_j \in s'$ there exist at least one $r_i \in r'$ so that $d_{s}(s_j, r_i) \leq \epsilon_{sp}$ and $d_{t}(s_j, r_i) \leq \epsilon_t$. \end{problem}

\begin{figure} [thb]
    \centering
    \includegraphics[width=.4\textwidth]{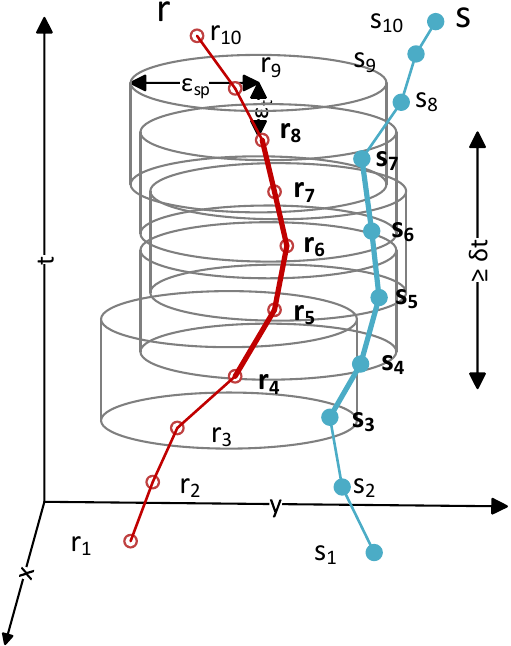}
    \caption{A pair of ``matching'' subtrajectories $(r_{4,8}, s_{3,7})$.}
  \label{fig_subtrajoin}
\end{figure}

Figure~\ref{fig_subtrajoin} illustrates two trajectories $r$ and $s$ and their respective \emph{matching} subtrajectories ($r_{4,8} ,s_{3,7}$). Each point of a trajectory defines a spatiotemporal 'neighborhood' area around it, i.e. a cylinder of radius $\epsilon_{sp}$ and height $\epsilon_t$. In order for a pair of subtrajectories to be considered matching, each point of a subtrajectory must have at least one point of the other subtrajectory in its neighborhood, thus making the result symmetrical. Furthermore the duration of the match should be at least $\delta t$.

The second step takes as input the result of the first step, which is actually a trajectory and neighboring trajectories and aims at segmenting each trajectory $r \in D$ into a set of subtrajectories. The way that a trajectory is segmented into subtrajectories is neighbourhood-aware, meaning that a trajectory will be segmented every time its neighbourhood changes significantly, so as to result in homogeneous subtrajectories (w.r.t. their surrounding moving objects). Returning to Example~\ref{ex:intro}, trajectory $A\rightarrow{D}$ should be segmented to $A\rightarrow{O}$ and $O\rightarrow{D}$, since at $O$ the cardinality and the composition of its neighbourhood changes significantly. The problem of trajectory segmentation can now be formulated as follows.

\begin{problem} \label{prb_segm}
\textbf{(Trajectory Segmentation)}
Given a trajectory $r$, identify the set of timestamps $CP$ (cutting points), where the density (or alternatively the composition) of the neighborhood of $r$ changes significantly. Then according to $CP$, $r$ is partitioned to a set of subtrajectories $\{r'_1,\dots,r'_M\}$, where $M = |CP|+1$ is the number of subtrajectories for a given trajectory $r$, such that $r = \bigcup_{k=1}^{M} r'_k$ and $k \in [1,M]$. 
\end{problem}

Given the output of Problem~\ref{prb_trj_join}, applying a trajectory segmentation algorithm for the trajectories $D$ will result in a new set of subtrajectories $D'$.

The third step takes as input $D'$ and the goal is to create clusters (whose cardinality is unknown) of similar subtrajectories and at the same time identify subtrajectories that are significantly dissimilar from the others (outliers). 
More specifically, let $C = \{C_1, \dots, C_K\}$ denote the clustering, where $K$ is the number of clusters, and for every pair of clusters $C_i$ and $C_j$, with $i,j \in [1,K]$, it holds that $C_i \cap C_j =  \text{\O}$.
Now, let us assume that each cluster $C_i \in C$ is represented by one subtrajectory, called \textit{Representative}, denoted as $R_i$. Actually, the problem of clustering is to discover clusters of objects such that the intra-cluster similarity is maximized and the inter-cluster similarity is minimized.
Therefore, the problem of subtrajectory clustering can be formulated as an optimization problem as follows.

\begin{problem} \label{prb_st_clust}
\textbf{(Subtrajectory Clustering and Outlier Detection)}
Given a set of subtrajectories $D'$, partition $D'$ into a set of clusters $C$ and a set of outliers $O$, where $D' = C \cup O$, in such a way so that the Sum of Similarity between Cluster members and cluster Representatives (\emph{SSCR}) is maximized: 
\begin{equation} \label{eqn:sscr}
SSCR = \sum\limits_{\forall R_i \in R}\sum\limits_{\forall r'_j \in C_i} Sim(R_i, r'_j)
\end{equation}
\end{problem}

However, trying to solve Problem \ref{prb_st_clust} by maximizing Equation (\ref{eqn:sscr}) is not trivial, since the problem to segment trajectories to subtrajectories, select the set of representatives $R$ and its cardinality $|R|$ that maximizes Equation (\ref{eqn:sscr}), has combinatorial complexity.

\subsection{Distributed Subtrajectory Clustering} \label{sec_distr_prob}

In this paper, we address the challenging problem of subtrajectory clustering in a distributed setting, where the dataset $D$ is stored distributed in different nodes, and centralized processing is prohibitively expensive.  

\begin{problem} \label{prb_distr_stclust}
\textbf{(\prob)}
Given a distributed set of trajectories, $D=\cup_{i=1}^{P} D_i$, where $P$ is the number of partitions of $D$, perform the subtrajectory clustering task in a parallel manner.
\end{problem}

Actually, Problem \ref{prb_distr_stclust} can be broken down to solving Problems \ref{prb_trj_join}, \ref{prb_segm} and \ref{prb_st_clust} (in that order) in a parallel/distributed way. 
In the following, we adopt this approach and outline a solution that is based on MapReduce.

\section{Problem Solution} \label{sec_sol}

\subsection{Overview} \label{sec_overview}
An overview of our approach is presented in Algorithm~\ref{alg_overv}. Initially, we \emph{Repartition} the data into $P$ equi-sized, temporally-sorted partitions (files), which are going to be used as input for the join algorithm in order to perform the subtrajectory join in a distributed way (line~\ref{lin:over_1}). 
Note that this is actually a preprocessing step that only needs to take place once for each dataset $D$. However, it is essential as it enables load balancing, by addressing the issue of temporal skewness in the input data. 
Subsequently, for each partition $D_i \in \cup_{i=1}^{P} D_i$ and for each trajectory we discover parts of other trajectories that moved close enough in space an time (line~\ref{lin:over_2}). Successively, we group by trajectory in order to perform the subtrajectory join (line~\ref{lin:over_3}). At this phase, since our data is already grouped by trajectory, we also perform trajectory segmentation in order to split each trajectory to subtrajectories (line~\ref{lin:over_4}). In turn, we utilize the temporal partitions created during the \emph{Repartition} phase and re-group the data by temporal partition. For each $D_i \in \cup_{i=1}^{P} D_i$ we calculate the similarity between subtrajectories and perform the clustering procedure (line~\ref{lin:over_5}). At this point we should mention that if a subtrajectory intersects the borders of two partitions, then it is replicated in both of them. This will result in having duplicate and possibly contradicting results. For this reason, as a final step, we treat this case by utilizing the \emph{Refine Results} procedure (line~\ref{lin:over_6}). Finally, a set $C$ of clusters and a set $O$ of outliers are produced. 

\begin{figure*} [t]
  \begin{center}
  \includegraphics[width=1\textwidth]{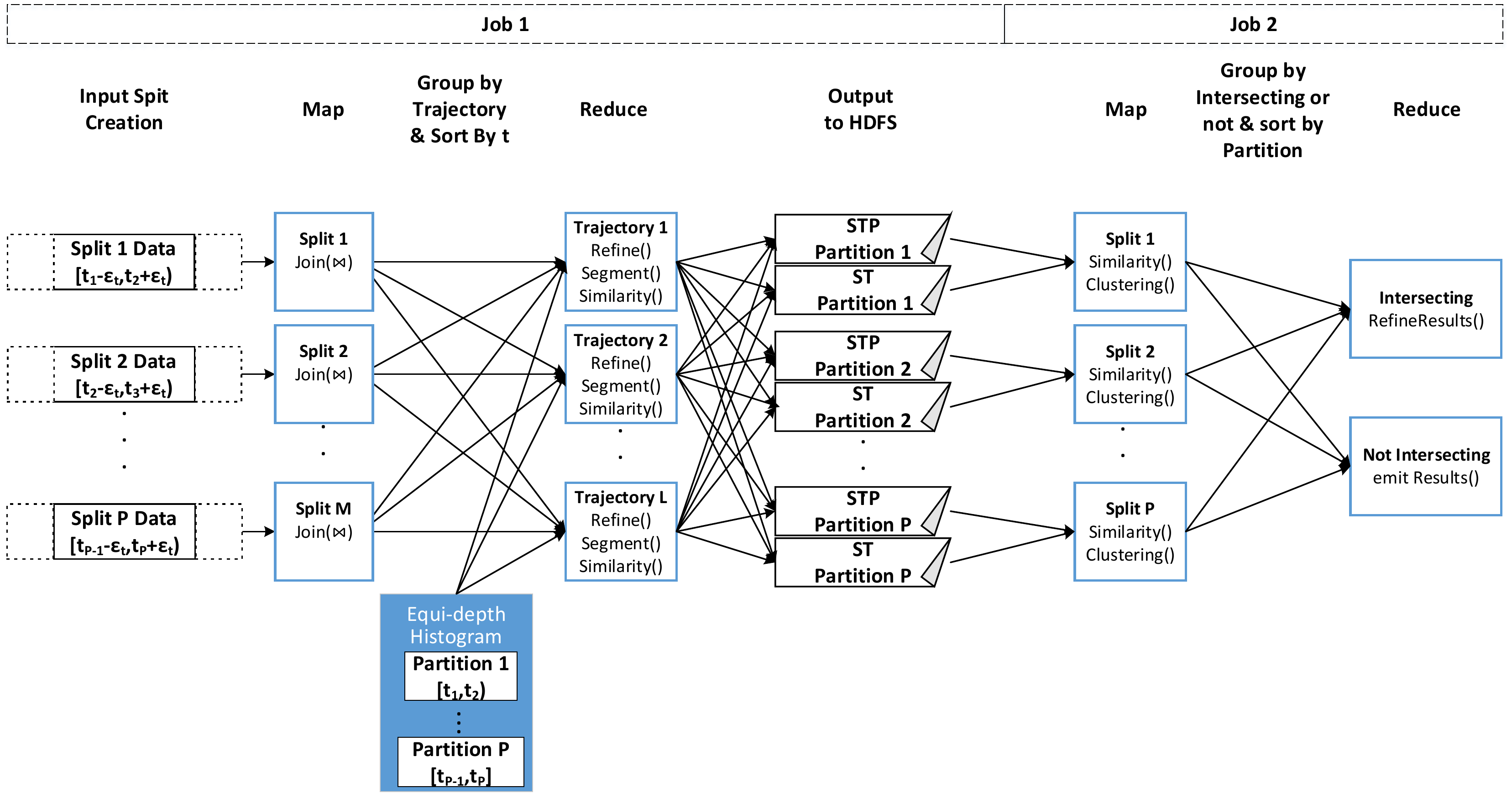}
  \caption{The \emph{DSC} algorithm. (Job 1) \emph{DTJ} and \emph{Trajectory Segmentation} and (Job 2) \emph{Clustering} and \emph{Refine Results}.}
  \label{fig_DSC}
  \end{center}
\end{figure*}

\begin{algorithm}[t]
\caption{$DSC$($D$)}
\label{alg_overv}
\begin{algorithmic}[1]
\STATE{\textbf{Input:} $D$}
\STATE{\textbf{Output:} set $C$ of clusters, set $O$ of outliers}
\STATE {\textbf{Preprocessing:} \emph{Repartition D};}\label{lin:over_1}
\FOR{\textbf{each} partition $D_i \in \cup_{i=1}^{P} D_i$}
    \STATE{\textbf{perform} \emph{Point-level Join};}\label{lin:over_2}
\ENDFOR
\STATE {\textbf{group by} \emph{Trajectory};}
\FOR{\textbf{each} Trajectory $r \in D$}
    \STATE{\textbf{perform} \emph{Subtrajectory Join}; -- \textit{Sect.~\ref{sec_dtj_sol}}}\label{lin:over_3}
    \STATE{\textbf{perform} \emph{Trajectory Segmentation};  -- \textit{Sect.~\ref{sec_dts_sol}}}\label{lin:over_4}
\ENDFOR
\STATE {\textbf{group by} $D_i$;}
\FOR{\textbf{each} subtrajectory $r' \in D_i$}
    \STATE{\textbf{calculate} $Similarity$ with other subtrajectories; -- \textit{Sect.~\ref{sec_dts_sol}}}\label{lin:over_5}
\ENDFOR
\STATE{\textbf{perform} $Clustering$; -- \textit{Sect.~\ref{sec_dc_sol}}}
\STATE{\textbf{perform} \emph{Refine Results};}\label{lin:over_6}
\STATE{\textbf{return} $C$ and $O$;}
\end{algorithmic}
\end{algorithm}

\subsection{Distributed Subtrajectory Join} \label{sec_dtj_sol}

As already mentioned, the first step is to perform the subtrajectory join in a distributed way. For this reason, we exploit the work presented   in~\cite{DBLP_journals/corr/abs-1903-07748}, coined \emph{DTJ}, which introduces an efficient and highly scalable approach to deal with Problem \ref{prb_trj_join}, by means of MapReduce. More specifically, \emph{DTJ} is comprised of a \emph{Repartitioning} phase and a \emph{Query} phase.

The \emph{Repartitioning} phase is a preprocessing step that takes place only once and it is independent of the actual parameters of the problem, namely $\epsilon_{sp}$, $\epsilon_t$, and $\delta t$. The idea is to construct an equi-depth histogram based on the temporal dimension, where each of the $M$ bins contain the same number of points and the borders of each bin correspond to a temporal interval $[t_i,t_j)$. The histogram is constructed by taking a sample of the input data\footnote{In Hadoop, this is achieved using the \emph{InputSampler} and \emph{TotalOrderPartitioner}.}. Then, the input data is partitioned to processing tasks based on the temporal intervals of the histogram bins. This guarantees temporal locality in each partition, as well as equi-sized partitions, thus balancing the load fairly. 

In the \emph{Query} phase, the actual join processing takes place. It consists of two steps, the \emph{Join} and the \emph{Refine} step, which are implemented as a \emph{Map} and a \emph{Reduce} function respectively. The output of this MapReduce job is for each trajectory $r \in D$ all the moving objects, with their respective portion of movement, that moved close enough in space and time for at least some time duration. 
In more detail, the output of \emph{DTJ} is per trajectory and the tuples are of the form $<refTrajPoint_{i},\{MatchingPoints\}>$, where $refTrajPoint_{i}$ is the i-th point of the reference trajectory, with $i \in [1, N]$ 
and $MatchingPoints$ is a list of points of other trajectories that have been identified as join results by the \emph{DTJ} query. 
In Figure~\ref{fig_DSC}, the \emph{DTJ} query corresponds to Job 1 until the Refine() procedure. 

For more technical details about the algorithms involved in \emph{DTJ} and an extensive experimental study, we refer to~\cite{DBLP_journals/corr/abs-1903-07748}.

\subsection{Distributed Trajectory Segmentation} \label{sec_dts_sol}

The \emph{Trajectory Segmentation} algorithm (TSA) takes as input a single trajectory, along with information about its neighborhood, and partitions it to a set of subtrajectories. In this paper, we propose two alternative segmentation algorithms.

The first algorithm, coined $TSA_{1}$, identifies the beginning of a new subtrajectory whenever the density of its neighborhood changes significantly. Such a segmentation algorithm is reminiscent of the flock definition~\cite{DBLP_journals/gis/LaubeIW05}, where the identified groups need to be composed of at least $m$ objects. For this purpose, we use the concept of \emph{voting} as a measure of density of the surrounding area of a trajectory. 
For a given point $r_i$ and any trajectory $s$, the voting $V(r_i)$ is defined as:

\begin{equation} \label{eqn:vot_point}
V(r_i) = \sum\limits_{\forall s \in D}\frac{d_{s}(r_i, s_k)}{\epsilon_{sp}}
\end{equation}

\noindent where, $s_k$ is the matching point of $s$ with $r_i$, as emitted by the subtrajectory join procedure. For a trajectory $r$ that consists of $N$ points $\{r_1,\dots,r_N\}$, we compute its normalized voting vector $\overline{V}(r)$ as follows:
\begin{equation} \label{eqn:vot_vector}
\overline{V}(r)[] = \{\frac{V(r_1)}{\max_{i=1}^{N} V(r_i)},\dots,\frac{V(r_N)}{\max_{i=1}^{N} V(r_i)}\}
\end{equation}

\noindent Finally, the voting of a trajectory (or subtrajectory) is defined as:
\begin{equation} \label{eqn:vot_traj}
V(r) = \frac{1}{N}\sum\limits_{i=1}^{N}V(r_i)
\end{equation}

The second segmentation algorithm, coined $TSA_{2}$, identifies the beginning of a new subtrajectory whenever the composition of its neighborhood changes substantially. This segmentation algorithm is reminiscent of the moving cluster definition~\cite{DBLP_conf/ssd/KalnisMB05}, where the identified groups need to share a sufficient number of common objects. Such an algorithm does not take as input the $\overline{V}(r)[]$ but instead, for each point $r_i \in r$, it takes as input a list $L(r_i)[]$ of the trajectory ids that have been produced as output by the \emph{DTJ} procedure. 

\begin{figure}[t]
\centering
\begin{subfigure}[thb]{0.25\columnwidth}
  \includegraphics[width=\columnwidth]{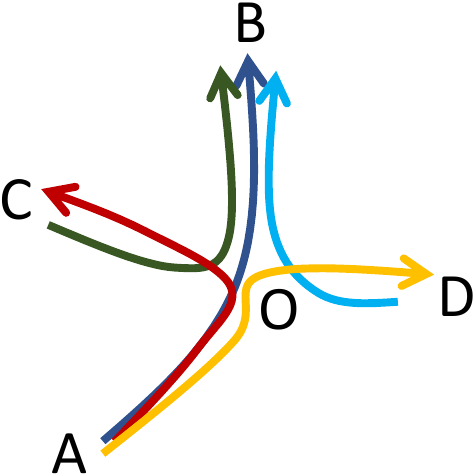}
  \caption{}
\end{subfigure}
~
\begin{subfigure}[thb]{0.25\columnwidth}
  \includegraphics[width=\columnwidth]{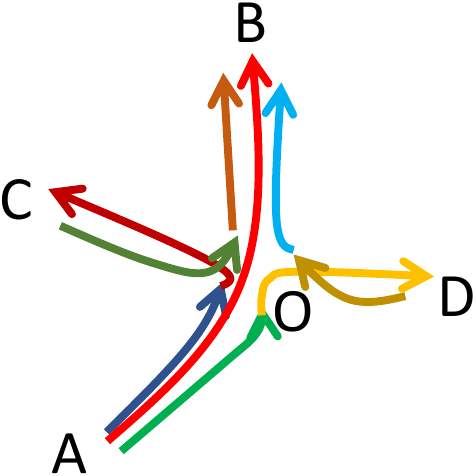}
    \caption{}
\end{subfigure}
~
\begin{subfigure}[thb]{0.25\columnwidth}
  \includegraphics[width=\columnwidth]{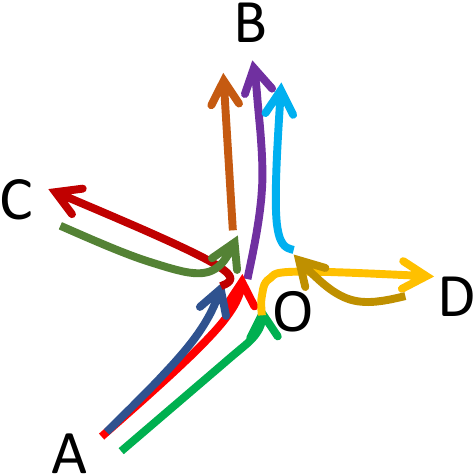}
    \caption{}
\end{subfigure}
\caption{(a) Five trajectories $A\rightarrow{B}$, $A\rightarrow{C}$, $A\rightarrow{D}$, $C\rightarrow{B}$ and $D\rightarrow{B}$, (b) $TSA_1$ segmentation, (c) $TSA_2$ segmentation}
\label{fig_tsa_1_2}
\end{figure}

The following example explains intuitively the difference between the two segmentation algorithms.

\begin{example} \label{ex:tsa} 
Consider the example of Figure~\ref{fig_tsa_1_2}(a) that illustrates five trajectories: $A\rightarrow{B}$, $A\rightarrow{C}$, $A\rightarrow{D}$, $C\rightarrow{B}$ and $D\rightarrow{B}$. Figures~\ref{fig_tsa_1_2}(b) and (c) depict the result of $TSA_1$ and $TSA_2$, respectively. In more detail, we can observe that both $TSA_1$ and $TSA_2$ segmented trajectory $A\rightarrow{D}$ to subtrajectories $A\rightarrow{O}$ and $O\rightarrow{D}$, due to the fact that after $O$, both the density and the composition of the neighborhood changes. The same holds for trajectories $A\rightarrow{C}$, $C\rightarrow{B}$ and $D\rightarrow{B}$, which are segmented to subtrajectories $A\rightarrow{O}$, $O\rightarrow{C}$, $C\rightarrow{O}$, $O\rightarrow{B}$, $D\rightarrow{O}$ and $O\rightarrow{B}$. However, when it comes to trajectory $A\rightarrow{B}$, we can observe that while $TSA_2$ segments it to subtrajectories $A\rightarrow{O}$ and $O\rightarrow{B}$, $TSA_1$ does not perform any segmentation. This is due to the fact that, after $O$, even though the density of the neighborhood remains the same (i.e. 3 moving objects), the composition of the neighborhood changes completely. In a subsequent step this will drive the clustering algorithm to identify, in the case of $TSA_1$ a flock-like cluster from $A$ to $B$, while in the case of $TSA_2$ two moving clusters from $A$ to $O$ and from $O$ to $B$. 
\end{example}

\begin{figure} [t]
  \begin{center}
  \includegraphics[width=0.48\textwidth]{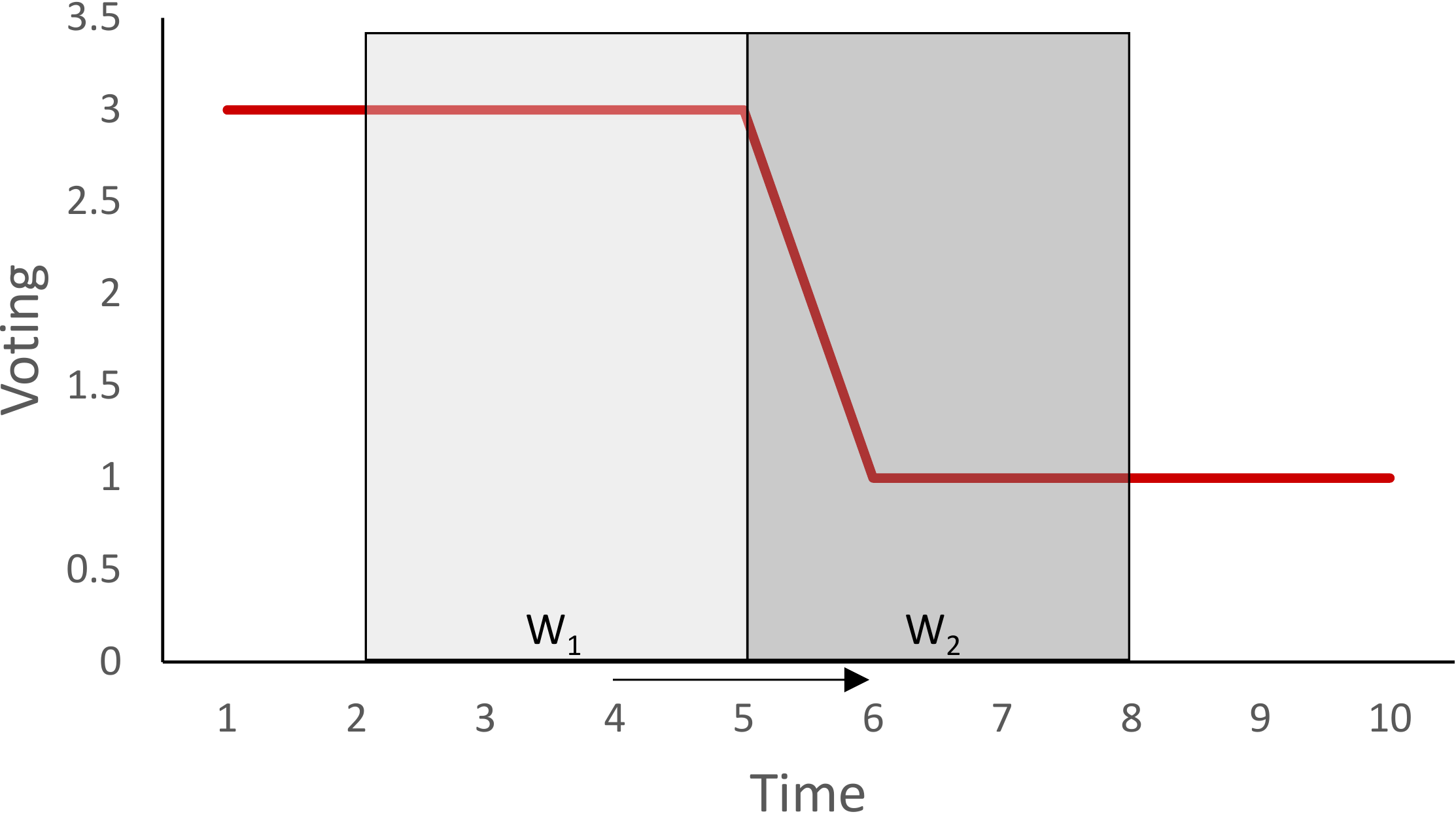}
  \caption{The two consecutive sliding windows $W_1$ and $W_2$ used by the segmentation algorithms.}
  \label{fig_tsa_w}
  \end{center}
\end{figure}

Both segmentation algorithms share a common methodology, which employs two consecutive sliding windows $W_1$ and $W_2$ of size $w$ (i.e. $w$ samples) to estimate the point $r_i \in CP$ (cutting point) where the ``difference'' between the two windows is maximized. This methodology has been successfully applied in the past on signal segmentation~\cite{DBLP_journals/tmm/PanagiotakisT05, DBLP_journals/tgrs/PanagiotakisKV08}. To exemplify, let us consider trajectory $A\rightarrow{D}$ of Example~\ref{ex:tsa}. For simplicity, we assume that the voting of the specific trajectory from $A$ to $O$ is 3 and from $O$ to $D$ is 1. Figure~\ref{fig_tsa_w} illustrates the two sliding windows $W_1$ and $W_2$ that traverse the voting signal of trajectory $A\rightarrow{D}$.

\textbf{Trajectory segmentation.}
Since the output of the \emph{DTJ} algorithm is per trajectory, it is straightforward to give it as input to \emph{TSA} which operates at the level of a trajectory. Moreover, the segmentation is performed in an embarrassingly parallel way, due to the fact that each trajectory can be processed by a different reduce task independently from others, as depicted in Figure~\ref{fig_DSC}. 
In more detail, for a given trajectory $r \in D$, $TSA_{1}$ first calculates the normalized voting vector $\overline{V}(r)[]$ and then performs the segmentation by utilizing it. Apart from $\overline{V}(r)[]$, the input of the \emph{TSA} algorithm is two additional parameters: $w$ and $\tau$. The output is a vector $CP[]$, which keeps the starting position of each subtrajectory of $r$. 

\begin{algorithm}[thb]
\caption{$TSA_{1}$($\overline{V}(r)[], w, \tau$)}
\label{alg_tsa1}
\begin{algorithmic}[1]
\STATE{\textbf{Input:} $\overline{V}(r)[], w, \tau$}
\STATE{\textbf{Output:} $CP[]$} 
\STATE $1 \rightarrow CP[]$;
\FOR{n = w+1 \dots N-w-1} \label{lin:1}
\STATE $m_1 = \frac{1}{w}\sum_{i=n-w}^{n-1} \overline{V}(r)[i]$;\label{lin:m1}
\STATE $m_2 = \frac{1}{w}\sum_{i=n}^{n+w-1} \overline{V}(r)[i]$;\label{lin:m2}
\STATE $d[n] = |m_1 - m_2|$;\label{lin:d}
\STATE $d_{max} = \max_{i = w+1}^{N-w-1} d[i]$;\label{lin:max}
\IF{$d[n] > \tau \wedge d[n] >= d_{max}$}\label{lin:traj_break1}
\STATE $n \rightarrow CP[]$;
\ENDIF\label{lin:traj_break2}
\ENDFOR
\end{algorithmic}
\end{algorithm}

In more detail, as presented in Algorithm~\ref{alg_tsa1}, two consecutive sliding windows of size $w$ are created over $\overline{V}(r)[]$, named $W_1$ and $W_2$ (line~\ref{lin:1}). These sliding windows move forward in time until $\overline{V}(r)[]$ is traversed. Here, $N$ is the number of points of trajectory $r \in D$. Then, for each window, the average normalized voting is computed (lines~\ref{lin:m1}-\ref{lin:m2}) and their absolute difference is stored in $d[]$, which is an array that stores all the differences between the sliding windows (line~\ref{lin:d}). Subsequently, we examine whether the current difference $d[n]$ is larger than the maximum difference $d_{max}$ and we update $d_{max}$ accordingly (line~\ref{lin:max}). Finally, if the difference $d[n]$ is higher than a threshold $\tau$ and is locally maximized, then, at that point, we segment the trajectory and we store the starting position of the new subtrajectory to $CP[]$ (lines~\ref{lin:traj_break1}-\ref{lin:traj_break2}).

On the other hand, the input of $TSA_{2}$ is a list of lists $L(r)[]$ for each $r \in D$. Similarly, two consecutive sliding windows $W_1$ and $W_2$ of size $w$ are created (line~\ref{lin:1_2}). 
Then, for each window, the union of lists is computed and stored in $l_1$ and $l_2$, respectively (lines~\ref{lin:l1}-\ref{lin:l2}). Successively, the Jaccard dissimilarity between $l_1$ and $l_2$ is computed and is stored to $d[]$, which is an array that stores all the similarities between the sliding windows (line~\ref{lin:d_2}). From then on, the algorithm is identical to $TSA_{1}$.

\begin{algorithm}[thb]
\caption{$TSA_{2}$($L(r), w, \tau$)}
\label{alg_tsa2}
\begin{algorithmic}[1]
\STATE{\textbf{Input:} $L(r), w, \tau$}
\STATE{\textbf{Output:} $CP[]$} 
\STATE $1 \rightarrow CP[]$;
\FOR{n = w+1 \dots N-w-1} \label{lin:1_2}
\STATE $l_1 = \cup_{i=n-w}^{n-1} L(r_i)[]$;\label{lin:l1}
\STATE $l_2 = \cup_{i=n}^{n+w-1} L(r_i)[]$;\label{lin:l2}
\STATE $d[n] = 1-\frac{|l_{1} \cap l_{2}|}{|l_{1}| + |l_{2}| - |l_{1} \cap l_{2}|}$;\label{lin:d_2}
\STATE $d_{max} = \max_{i = w+1}^{N-w-1} d[i]$;\label{lin:max_2}
\IF{$d[n] > \tau \wedge d[n] >= d_{max}$}\label{lin:traj_break1_2}
\STATE $n \rightarrow CP[]$;
\ENDIF\label{lin:traj_break2_2}
\ENDFOR
\end{algorithmic}
\end{algorithm}


\textbf{Similar subtrajectories.}
After trajectory segmentation, the next step is to calculate the similarity between all the pairs of subtrajectories, using Equation~\ref{eqn:sim_traj}. This cannot be done completely after the segmentation at the \emph{Reducer} phase of Job 1, illustrated in Figure~\ref{fig_DSC}, because at that point each reduce function has information only about the segmentation of the reference trajectory to subtrajectories. For this reason, at this point we cannot calculate the denominator of Equation~\ref{eqn:sim_traj}. However, for each subtrajectory $r' \in r$, where $r$ is the reference trajectory, we can calculate the similarity between the matching points (enumerator of Equation~\ref{eqn:sim_traj}). 

In more detail the output of each reduce function (Job 1 Figure~\ref{fig_DSC}) is a relation, called \emph{STP}, which holds a set of key-value pairs of the form $<(r'.ID, s.ID), \{(s_{f}.t, Sim(s_{f}, r'))$ $\dots$ $(s_{l}.t, Sim(s_{l}, r')) \}>$, where $s_{f},s_{l}$ are the temporal first and last point, respectively, of trajectory $s$ that ``matches'' with subtrajectory $r'$. Moreover, in a separate relation, coined \emph{ST}, we hold some extra information for each subtrajectory. More specifically, the tuples of $ST$ are key-value pairs, where the key is the subtrajectory identifier $<ID>$ and the value is of the form  $<t_{s}, t_{e}, V, Card>$, where $t_s$ ($t_e$) is the starting time (ending time, respectively) of the subtrajectory, $V$ is the voting and $Card$ is the number of points which constitute the specific subtrajectory. Due to the fact that these two relations can be pretty large, we need to partition them into smaller files. In order to achieve this, we broadcast the load balanced temporal partitions that were created during the \emph{Repartitioning} phase of \emph{DTJ}. As illustrated in Figure~\ref{fig_DSC}, each reducer loads these partitions and assigns each subtrajectory (tuple of \emph{ST} and \emph{STP}) to all the partitions with which it temporally intersects. Subsequently, the tuples are grouped by temporal partition and each group is fed to a Mapper.

At this point, each \emph{Mapper} has now all the information needed to calculate the similarity between all the pairs of subtrajectories (Equation~\ref{eqn:sim_traj}), for each temporal partition separately. The similarity between subtrajectories is output in a new relation, called \emph{SP}. Each tuple of this relation holds information about a subtrajectory $r'$ and its similarity with all the other subtrajectories, whenever this similarity is larger than zero. More specifically, \emph{SP} contains a set of key-value pairs where the key is the ID of the subtrajectory $(r'.ID)$ and the value is a list $AdjLst$ containing elements of the form $(s'.ID, Sim)$, where $s'$ is a subtrajectory for which it holds that $Sim(r',s')>0$.

\subsection{Distributed Clustering} \label{sec_dc_sol}

\textbf{Clustering.} 
After having calculated the similarity between all pairs of subtrajectories for each temporal partition, we can proceed to the actual clustering and outlier detection procedure. The output of the similarity calculation process, namely $SP$, is actually an adjacency list. The intuition behind the proposed solution to Problem \ref{prb_st_clust} is to select as cluster representatives, highly voted subtrajectories (Equation~\ref{eqn:vot_traj}) that are not similar with the already selected representatives $R_i \in R$. Then, we assign each subtrajectory $r'_k$ to the $R_i$ (and hence $C_i$) with which it has the maximum similarity $Sim(r'_k, R_i)$. 

The input of the clustering algorithm is $SP$, $ST$ and parameters $k$ and $\alpha$ and the output is the set of clusters $C$ and the set of outliers $O$. More specifically, $k$ is a threshold for setting a lower bound on the voting of a representative. This prevents the algorithm from identifying clusters with small support. Parameter $\alpha$ is a similarity threshold used to assign subtrajectories to cluster representatives. It ensures that a subtrajectory assigned to a cluster has sufficient similarity with the representative of the cluster. This actually poses a lower bound to the average distance between the representatives and the cluster members and, consequently, guarantees a minimum quality in the identified clusters (intra-cluster distance).

\begin{lemma} \label{lem_rmse}
The average distance $\overline{d_{s}(r',s')}$, between a representative subtrajectory $r'$ and a cluster member $s'$ will always be at most $\epsilon_{sp} \cdot (1 - \alpha)$.
\begin{equation} \label{eqn:avg_dist}
\overline{d_{s}(r',s')} \leq \epsilon_{sp} \cdot (1 - \alpha)
\end{equation}
\end{lemma}


\begin{algorithm}[t]
\caption{$Clustering$($SP, ST, k, \alpha$)}
\label{alg_clust}
\begin{algorithmic}[1]
\STATE{\textbf{Input:} $SP, ST, k, \alpha$}
\STATE{\textbf{Output:} set $C$ of clusters, set $O$ of outliers}
\STATE{\textbf{sort} $ST$ by $V$ in descending order;}\label{lin:sort}
\FOR{each element $st \in ST$}
    \IF{$st \not\in R$}\label{lin:check_c}
        \IF{$st.V \geq k$}\label{lin:repr1}
            \STATE $st \rightarrow R$;\label{lin:repr2}
            \FOR{each element $l \in st.AdjLst$}\label{lin:for_l}
                \IF{$l \not\in C$}\label{lin:c_l1}
                    \IF{$Sim(l,st) \geq \alpha$}
                        \STATE $l \rightarrow  C(st)$;
                        \IF{$l \in O$}
                            \STATE $O = O - l$;\label{lin:c_l2}
                        \ENDIF
                    \ELSE
                        \STATE $O = O \cup l$
                    \ENDIF
                \ELSE
                    \IF{$Sim(l,st) > Sim(l,R(l))$}\label{lin:c_l3}
                        \STATE $C(R(l)) = C(R(l)) - l$;
                        \STATE $l \rightarrow  C(st)$;\label{lin:c_l4}
                    \ENDIF
                \ENDIF
            \ENDFOR
        \ELSE
            \STATE $O = O \cup st$;\label{lin:c_out}
        \ENDIF
    \ENDIF
\ENDFOR
\STATE {$C = C \cup R$}\label{lin:concat}
\end{algorithmic}
\end{algorithm}

To begin with, we want to traverse the subtrajectories by their voting, in descending order (i.e. highly voted subtrajectories first). In order to achieve this, we need to sort $ST$ by $V$ (line~\ref{lin:sort}). Subsequently, for each subtrajectory $st \in ST$ we examine whether it is already assigned to cluster (line~\ref{lin:check_c}). 
If $st$ is not assigned to any cluster and the voting of $st$ is less than $k$, then we add $st$ to the outliers set (line~\ref{lin:c_out}). Otherwise, we create a new cluster and consider $st$ as the representative (lines~\ref{lin:repr1}-\ref{lin:repr2}). Successively, we consult relation $SP$ and retrieve the adjacency list of $st$ (line~\ref{lin:for_l}). Then, for each element $l$ that belongs to the adjacency list of $st$, we examine if it is assigned to any cluster. If not, we investigate whether the similarity between $l$ and $st$ is greater or equal than the similarity threshold $\alpha$. If not, we add $l$ to the outlier set $O$, otherwise we assign it to the cluster led by $st$ and remove it from the outliers $O$, in case $l \in O$ (lines~\ref{lin:c_l1}-\ref{lin:c_l2}). If $l$ is assigned to a cluster, we examine whether the similarity of $l$ with $st$ is greater than the similarity with the representative of the cluster that $l$ is currently assigned. If this is the case, then we remove $l$ from the current cluster and assign it to the cluster led by $st$ (lines~\ref{lin:c_l3}-\ref{lin:c_l4}). Finally, we concatenate $C$ with $R$ (line~\ref{lin:concat}) so as to return, except from the outlier set $O$, both cluster members and representatives. 

\textbf{Refinement of Results.}  
At this point we successfully accomplished to deal with Problem \ref{prb_st_clust} for each temporal partition. However, this might result in having duplicates due to the fact that each subtrajectory that temporally intersects multiple partitions is replicated to each one of them. The actual problem that lies here is not the duplicate elimination problem itself but the fact that the result for such a subtrajectory might be contradicting in different partitions. In more detail, for each partition, the clustering procedure will decide whether a subtrajectory is a \emph{Representative} ($Repr$), a \emph{Cluster Member} ($Cl$) or an \emph{Outlier} ($Out$). Hence, for each intersecting subtrajectory $q$ and for each pair of consecutive partitions $(i,j)$ with which $q$ intersects, $q$ can have the following pairs of states: (a) $Out$-$Out$, (b) $Repr$-$Repr$, (c) $Cl$-$Cl$, (d) $Repr$-$Cl$ ($Cl$-$Repr$), (e) $Repr$-$O$ ($O$-$Repr$) and (f) $Cl$-$O$ ($O$-$Cl$). 

In order to implement the above procedure we need to have all the information concerning the intersecting subtrajectories ($C$ and $O$) for all the Partitions sorted in time. To do this, we group the trajectories according to whether they are intersecting or not. As illustrated in Figure~\ref{fig_DSC}, the non-intersecting are emitted, since they are not affected, while the intersecting subtrajectories get sorted by partition. Hence, a \emph{Reducer} will receive all the required information to make the appropriate decisions. In more detail, we sweep through the temporal dimension and for each pair of consecutive partitions we make the appropriate decisions.

\begin{algorithm}[t]
\caption{$RefineResults$($q$)}
\label{alg_clust_refine}
\begin{algorithmic}[1]
\STATE{\textbf{Input:} \emph{Intersecting Subtrajectories}}
\STATE{\textbf{Output:} set $C$ of clusters, set $O$ of outliers}
\FOR{each pair p $\rightarrow (P_{i},P_{i+1})$ of Partitions}
    \STATE{$P_{i} \cap P_{i+1} \rightarrow I$}
    \FOR{each element $e \in I$}
        \SWITCH {$p$}
            \CASE {(a)}
                \STATE \textbf{remove} $q$ from $O_{i}$;
            \ENDCASE
            \CASE {(b)}
                \STATE \textbf{merge} $C_{i}(q)$ and $C_{i+1}(q)$;
            \ENDCASE
            \CASE {(c)}
              \IF{$Sim(q, R_{i}(q)) > Sim(q, R_{i+1}(q))$}
                \STATE \textbf{remove} $q$ from $C_{i+1}$;
              \ELSE
                \STATE \textbf{remove} $q$ from $C_{i}$;
              \ENDIF
            \ENDCASE
            \CASE {(d)}
              \STATE \textbf{remove} $q$ from $C$;
            \ENDCASE
            \CASE {(e),(f)}
              \STATE \textbf{remove} $q$ from $O$;
            \ENDCASE
        \ENDSWITCH
    \ENDFOR
\ENDFOR
\end{algorithmic}
\end{algorithm}

For each of the above cases, as depicted in Algorithm~\ref{alg_clust_refine}, a decision has to be made, in order to eliminate duplicates and provide the correct result according to the problem definition. More specifically, in case of (a), $q$ is marked as outlier in both partitions, hence, we only need to eliminate duplicates. In case of (b), the two clusters are ``merged'', since all of the subtrajectories that belong to them are similar ``enough'' with $q$, which is the representative of both clusters. In case of (c), let us assume that $q$ belongs to cluster $C_i(R(q))$ in Partition $i$ and $C_{i+1}(R(q))$ in Partition $i+1$. Then, $q$ is assigned to the cluster with which it has the largest similarity with its representative. In case of (d), $q$ remains to be a cluster representative and is removed from the cluster $C$ in which it is a member. Finally, in case of (e) and (f), $q$ is removed from $O$.

\section{Complexity Analysis} \label{sec_complx}

The purpose of this section is to analyse and provide insight to the complexity of the different algorithms that are involved to the solution to the \emph{Distributed Subtrajectory Clustering} problem, presented in this paper. 

\textbf{\emph{DTJ}}: The complexity of the \emph{Join} algorithm is roughly $O(|D|log_2 Q)$, with $Q$ being the average number of points per spatial index partition and $Q << |D|$. The complexity of the \emph{Refine} algorithm is $O(T\cdot SW\cdot dt\cdot l)$, where $T$ is the average number of points per trajectory, $SW$ is the average number of points contained in a $\delta t + 2 \epsilon_t$ window, $dt$ the average number of points contained in a $\delta t$ window and $l$ is average the size of the $MatchingPoints$ list. For more details about the complexity of the algorithms involved in \emph{DTJ} please refer to~\cite{DBLP_journals/corr/abs-1903-07748}.

\textbf{\emph{Segmentation}}: The complexity of the \emph{$TSA_{1}$} algorithm is $O(l \cdot |T|)$, where $l$ is average the size of the ``matching'' list and $|T|$ is the average number of points per trajectory. The reason that we include $l$ to this analysis is that in order to perform \emph{$TSA_{1}$}, we first need to calculate the normalized voting vector. The complexity of the \emph{$TSA_{2}$} algorithm is also $O(l \cdot |T|)$, since $l_1$ and $l_2$ are already sorted by trajectory id and the list intersection can take place in linear time to the size of the lists.

\textbf{\emph{Clustering}}: The complexity of $Clustering$ algorithm is $O(|ST| \cdot log|ST| + |ST| \cdot |L|)$, with $|ST|$ being the number of subtrajectories, $|L|$ the average size of the adjacency list $AdjLst$ and $|ST| \cdot log|ST|$ is the sorting cost. Here, we should mention that $|ST| << |D|$. Furthermore, $ST$ and $SP$ are implemented as HashMaps, hence key search has an $O(1)$ time complexity.
The complexity of the $RefineResults$ algorithm is $O(M \cdot |P| \cdot |I|)$, where $M$ is the number of temporal partitions, $|P|$ is the average number of intersecting subtrajectories per partition and $I$ is the average size of the intersection. We should mention, here, that the intersection between two consecutive partitions is performed in linear time by utilizing HashSets sets.

\section{Experimental Study} \label{sec_exper}

In this section, we present the findings of our experimental evaluation. The experiments were conducted in a 49 node Hadoop 2.7.2 cluster,  provided by \emph{~okeanos}\footnote{IAAS service for the Greek Research and Academic Community \url{https://okeanos.grnet.gr/home/}}. The master node consists of 8 CPU cores, 8 GB of RAM and 60 GB of HDD while each slave node is comprised of 4 CPU cores, 4 GB of RAM and 60 GB of HDD. Our configuration enables each slave node to launch 4 containers, thus up to 192 tasks (\emph{Map} or \emph{Reduce}) can be launched simultaneously. 
For our experimental study, we employed two real datasets that will assist us to evaluate the performance, scalability and effectiveness of our solution. Furthermore, we utilized a synthetic dataset that simulates the case of Figure~\ref{fig_intro} in order to verify that our solution operates as anticipated, given a dataset with a known ground truth. The real datasets are from two different domains, namely the urban and the maritime domain. In more detail, the first one, named SIS\footnote{This private dataset was kindly provided by Gruppo Sistematica SpA}, is a 27GB proprietary insurance dataset of moving objects around Rome and Tuscany area, that contains approximately $2.2\times10^7$ trajectories that correspond to $7.2\times10^8$ points. The second one, coined Brest\footnote{\url{https://zenodo.org/record/1167595\#.XKHTyaRRVPa}}, is a 650MB publicly available AIS dataset of vessels moving in the wider Brest area, consisting approximately of $3.65\times10^5$ trajectories that correspond $17\times10^6$ points. 

Our experimental methodology is as follows: Initially, in Section~\ref{sec_comp} we verify the correctness of our solution by applying it to a dataset with a known ground truth and compare our findings with T-OPTICS~\cite{DBLP_journals/jiis/NanniP06}, a well-known entire trajectory clustering technique. Moreover, we compare our solution with TraClus~\cite{DBLP_conf/sigmod/LeeHW07} and S$^2$T-Clustering~\cite{DBLP_conf/edbt/PelekisTVPT17}, two state of the art subtrajectory clustering methods. Subsequently, in Section~\ref{sec_scalab}, we study the scalability of our solution by varying (a) the dataset size, and (b) the number of cluster nodes. Finally, in Section~\ref{sec_sensit}, we perform a sensitivity analysis in order to evaluate the effect of setting different values to the parameters of our solution, in terms of execution time and quality. Table~\ref{tab:params} shows the experimental setting, where we vary the following parameters: $\epsilon_{sp}$, $\epsilon_t$, $\delta t$, $w$, $\tau$, $\alpha$ and $k$ and measure their effect in the performance and the effectiveness of our algorithms. We should mention that the default segmentation algorithm in our experimental study is $TSA_1$.

\begin{table}[t]
\begin{small}
\begin{center}
\caption{Parameters and default values (in bold)} \label{tab:params}
\begin{tabular}{|c||l|l|l|l|l|} \hline
\multirow{1}{*}{\textbf{Parameter}} & \multicolumn{5}{c|}{\textbf{Values}} \\ \cline{2-6}
  & (i) & (ii) & (iii) & (iv) & (v) \\  \hline \hline
$\epsilon_{sp}$ (\%) & 10\% & 15\% & \textbf{20\%} & 25\% & 30\% \\ \hline
$\epsilon_t$ (\%)	& 0\% & 25\% & \textbf{50\%} & 75\% & 100\% \\ \hline
$\delta t$ (\%)	& 0\% & 25\% & \textbf{50\%} & 75\% & 100\% \\ \hline
$w$	&	10 & 15 & \textbf{20} & 25 & 30 \\ \hline
$\tau$	&	0.2 & 0.4 & \textbf{0.6} & 0.8 & 1 \\ \hline
$\alpha$ (in $\sigma$) & -2 & -1 & \textbf{0} & 1 & 2 \\ \hline
$k$ (in $\sigma$)&	-2 & -1 & \textbf{0} & 1 & 2 \\ \hline
\end{tabular}
\end{center}
\end{small}
\end{table}

\subsection{Parameter Setting} \label{sec_param_set}

Setting the different parameters for different datasets can turn out to be an arbitrary procedure, which, in turn, can jeopardise the quality of the clustering results. For this reason, we provide some simple rules for setting the parameters relatively to the dataset being clustered, that do not compromise the quality of the results. 
In more detail, $\epsilon_{sp}$ can be set as a percentage of the dataset diameter. This, however, can be problematic when dealing with datasets having large spatial variation in their density (e.g. ports in the maritime domain). For this reason, we utilized the partitioning provided by the spatial index (QuadTree) of \emph{DTJ} and calculated $\epsilon_{sp}$ for each point, as a percentage of the diameter of the cell of the QuadTree to which it belongs. Moreover, $\epsilon_{t}$ and $\delta t$ are calculated relatively to the average duration between two consecutive trajectory samples ($\approx$ 1200 sec for SIS and $\approx$ 950 sec for AIS Brest). 

Parameter $w$ sets the size of the windows $W_1$ and $W_2$ upon which some measure is calculated. Small values on $w$ can affect the robustness of the estimation, thus resulting to over-segmentation. On the other hand, large values of $w$ can result to overlooking some cutting points due to the large window size. It has been observed that for $w \approx 20$ the robustness of the estimation is not affected and the size of the window is small enough so as not to overlook any cutting points. Concerning parameter $\tau$, our experiments show that the best result in terms of quality is achieved for $\tau \approx 0.4$
Finally, the values of $\alpha$ and $k$ can be set ``around'' the mean value of the similarity and the voting of the temporal partition, respectively, in terms of standard deviation. In fact, it has been observed that the average similarity and voting can produce clustering results of good quality. For more details about the effect, in terms of quality, of setting different values to the parameters of our solution, please refer to Section~\ref{sec_sensit}

\begin{figure}[thb]
\centering
\begin{subfigure}[t]{0.48\columnwidth}
  \hspace*{-0.1cm}\includegraphics[width=\columnwidth]{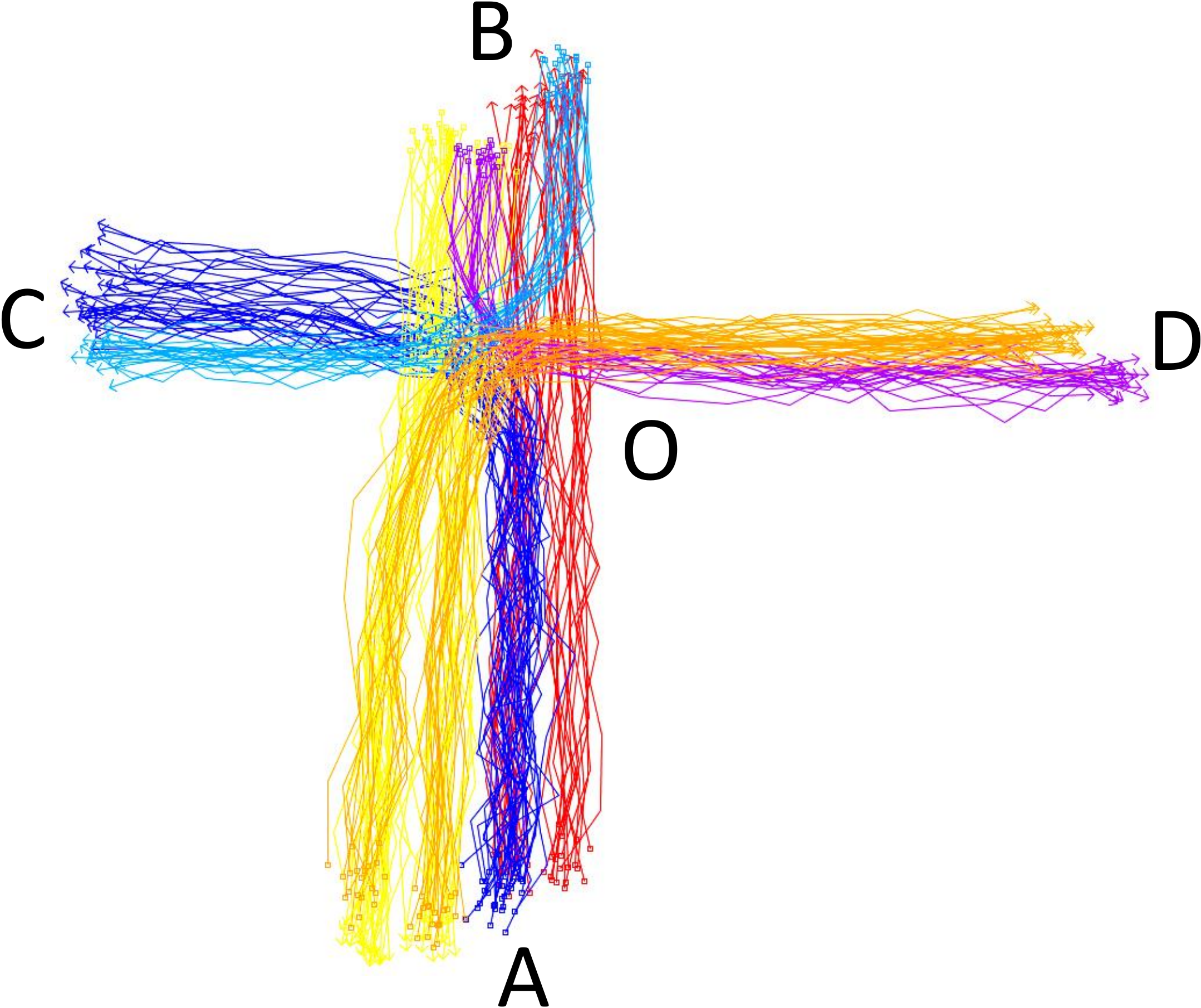}
  \caption{}
\end{subfigure}
~
\begin{subfigure}[t]{0.48\columnwidth}
  \includegraphics[width=\columnwidth]{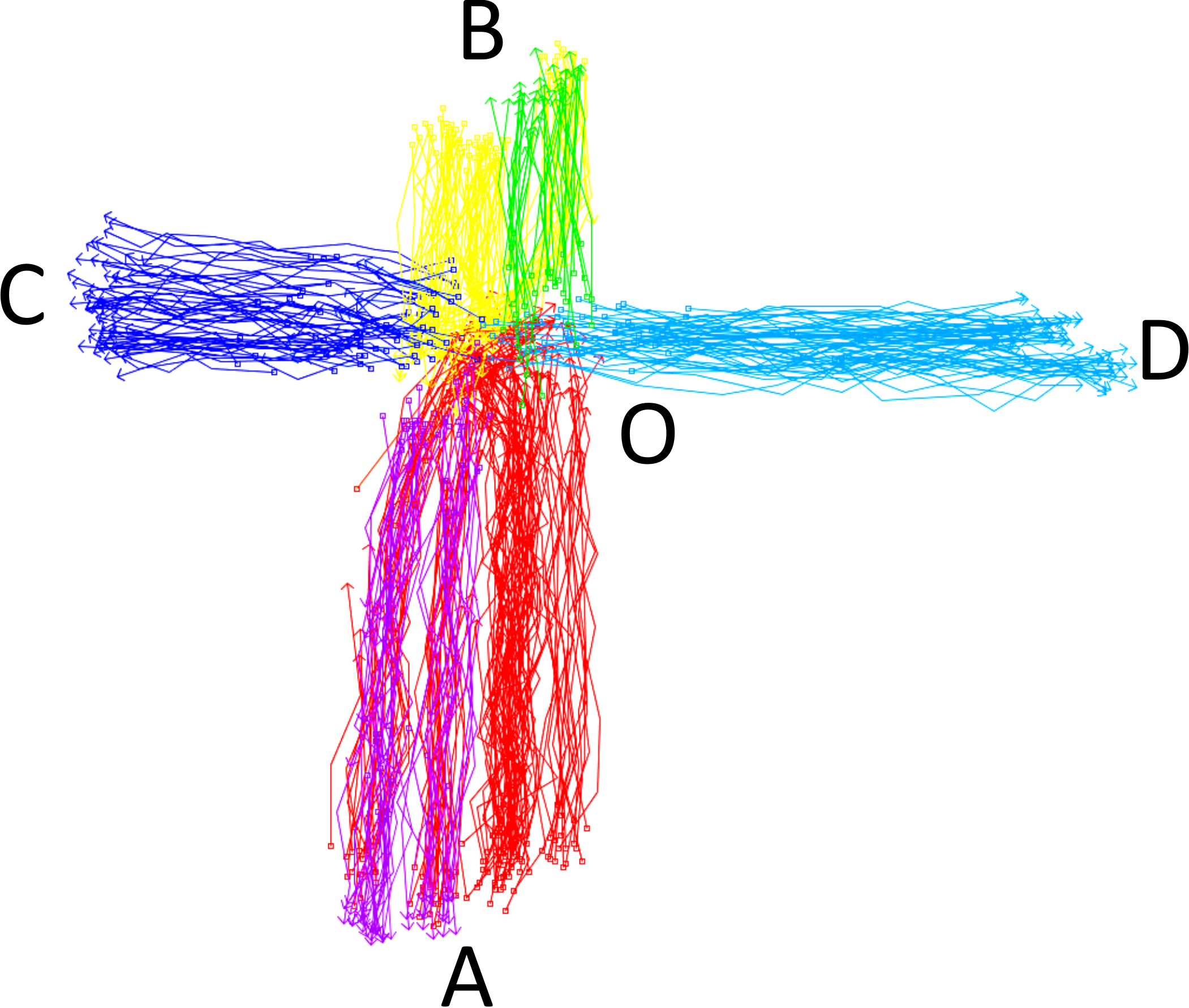}
    \caption{}
\end{subfigure}
\caption{Identified clusters by (b) \emph{T-OPTICS} and (b) \emph{DSC}}
\label{fig_groud_truth}
\end{figure}

\subsection{Comparison with related work} \label{sec_comp}

Initially, so as to verify that our solution operates as expected, we utilize a synthetic dataset\footnote{The original dataset was found in~\cite{DBLP_conf/cvpr/MorrisT09}} that simulates the case of Figure~\ref{fig_intro}. The only difference is that the two outliers mentioned there ($O\rightarrow{A}$ and $O\rightarrow{B}$), will now form clusters. Hence, the ground truth for the synthetic cluster becomes $A\rightarrow{O}$, $B\rightarrow{O}$, $O\rightarrow{C}$, $O\rightarrow{D}$, $O\rightarrow{A}$ and $O\rightarrow{B}$.

In fact, as depicted in Figure~\ref{fig_groud_truth}(a), T-OPTICS identifies the six original routes: $A\rightarrow{B}$ (in red), $A\rightarrow{C}$ (in blue), $A\rightarrow{D}$ (in orange), $B\rightarrow{A}$ (in yellow), $B\rightarrow{C}$ (in light blue) and $B\rightarrow{D}$ (in purple). On the other hand, \emph{DSC} identifies, with $Accuracy = 100\%$ and $F$-$measure = 1$, the six expected clusters of subtrajectories: $A\rightarrow{O}$(in red), $B\rightarrow{O}$ (in yellow), $O\rightarrow{C}$ (in blue), $O\rightarrow{D}$ (in light blue), $O\rightarrow{A}$ (in purple) and $O\rightarrow{B}$ (in green).

Subsequently, we compare \emph{DSC} with two state of the art subtrajectory clustering algorithms, \emph{S$^{2}$T-Clustering} and \emph{TraClus}. The metric that we employ in order to evaluate the quality of the outcome of the clustering procedure is the well-known \emph{RMSE} metric, which is actually a measure of intra-cluster distance between the representatives and the cluster members. Hence the larger the \emph{RMSE}, the lower the intra-cluster distance and consequently the quality of the clustering. It is obvious that, under this definition, \emph{RMSE} is equivalent to \emph{SSRC} (Equation~\ref{eqn:sscr}). In order to perform this experiment, we utilized the 20\% of each dataset which was further partitioned in 4 portions (25\%, 50\%, 75\%, 100\%). This choice was necessary because the centralized implementations of \emph{S$^{2}$T-Clustering} and \emph{TraClus} could not scale with the full size of the datasets that we utilized.

\begin{figure}[thb]
\centering
  \includegraphics[width=.45\columnwidth]{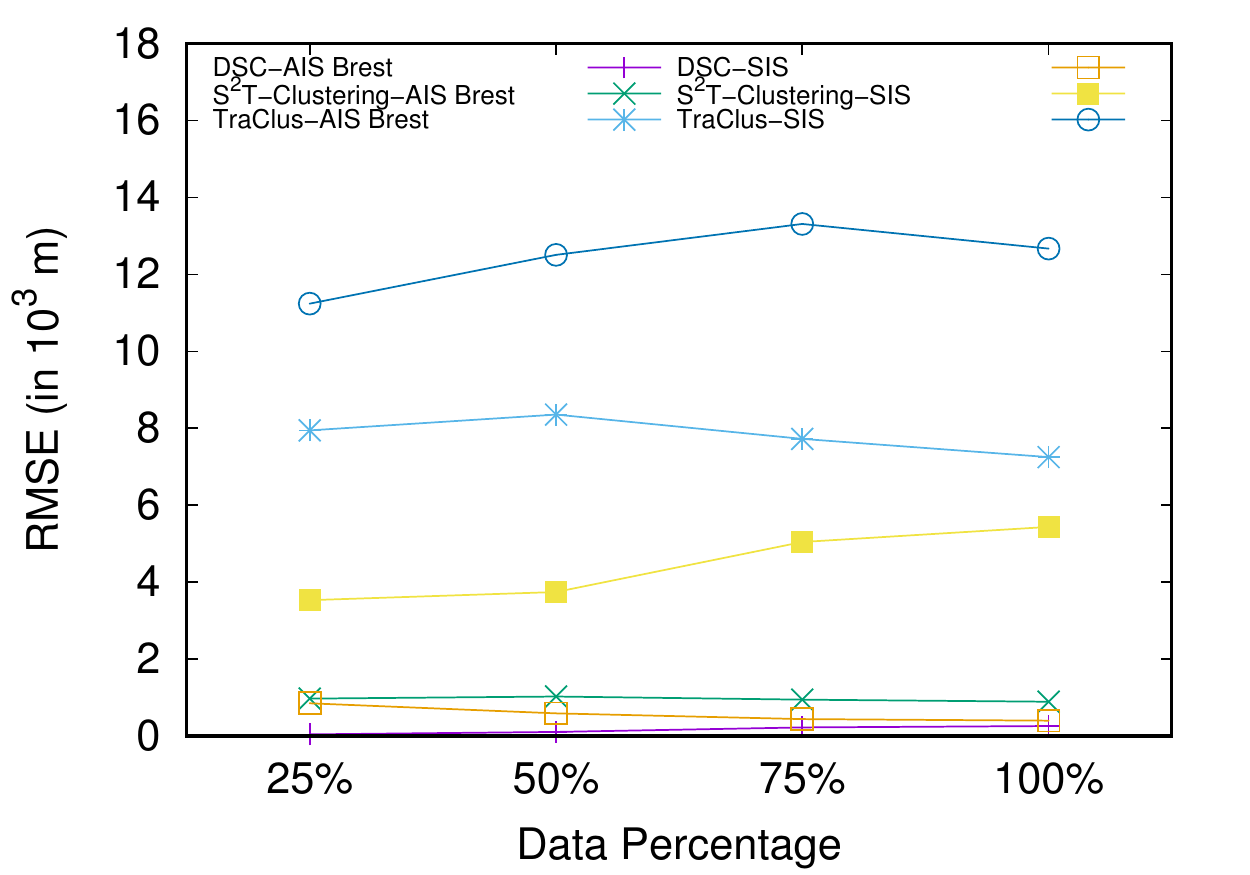}
  \caption{Comparison of the RMSE metric between \emph{DSC}, \emph{S$^{2}$T-Clustering} and \emph{TraClus}}
\label{fig_comp}
\end{figure}

As illustrated in Figure~\ref{fig_comp}, \emph{DSC} outperforms, in terms of \emph{RMSE}, both \emph{TraClus} and \emph{S$^{2}$T-Clustering}. In more detail, \emph{TraClus} presents the largest \emph{RMSE} which is somehow anticipated, since the specific algorithm utilizes a density-based approach to cluster subtrajectories, which in turn, through cluster expansion, can lead to spatially extended clusters. On the other hand, \emph{S$^{2}$T-Clustering} presents smaller \emph{RMSE} than \emph{TraClus}, due to the fact that it adopts a distance-based approach and discovers more compact clusters. However, \emph{DSC} results in smaller \emph{RMSE} than \emph{S$^{2}$T-Clustering}, mostly due to the fact that in the latter, two trajectories might end-up in the same cluster even if they have small ``matching portions''. However, in \emph{DSC} this ``matching portions'' should have a minimum ($\delta t$) duration.

\begin{figure*}[thb]
\centering
\begin{subfigure}[t]{0.48\columnwidth}
  \includegraphics[width=\columnwidth]{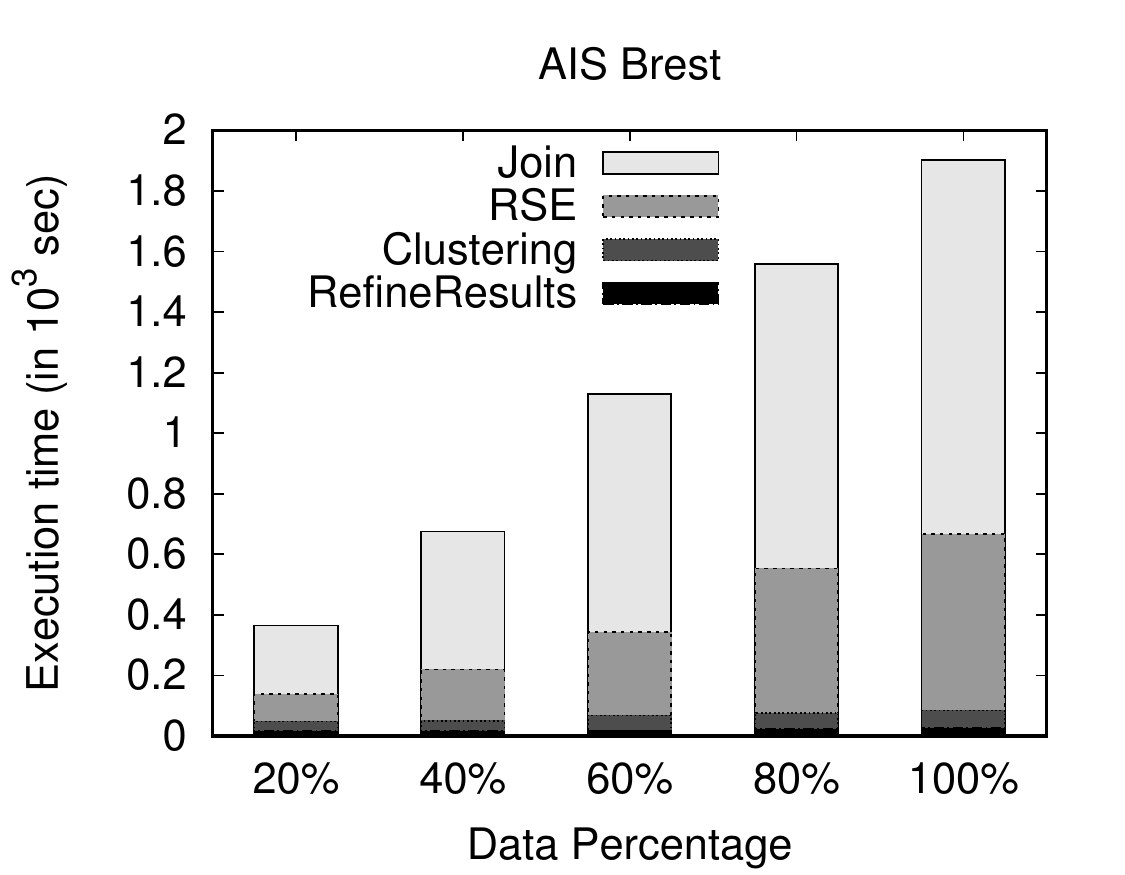}
  \caption{}
\end{subfigure}
~
\begin{subfigure}[t]{0.48\columnwidth}
  \includegraphics[width=\columnwidth]{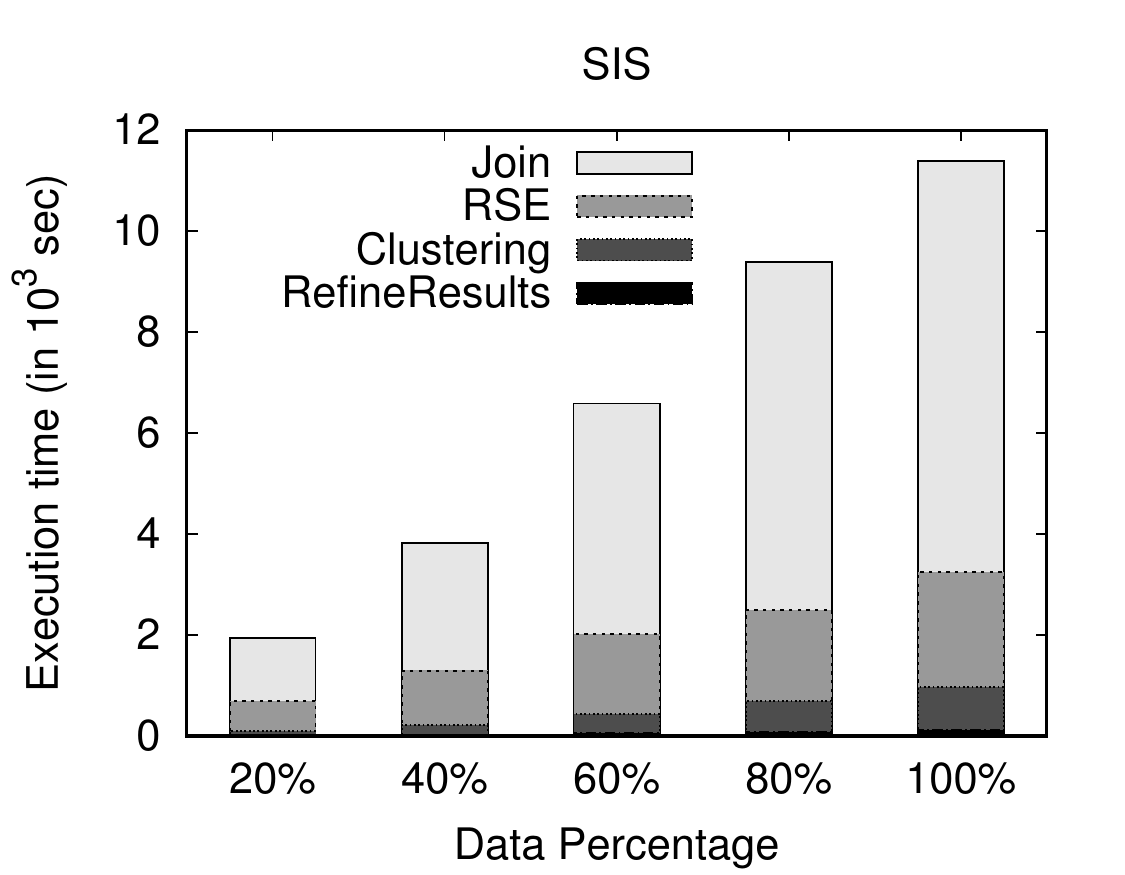}
  \caption{}
\end{subfigure}
~
\begin{subfigure}[t]{0.48\columnwidth}
  \includegraphics[width=\columnwidth]{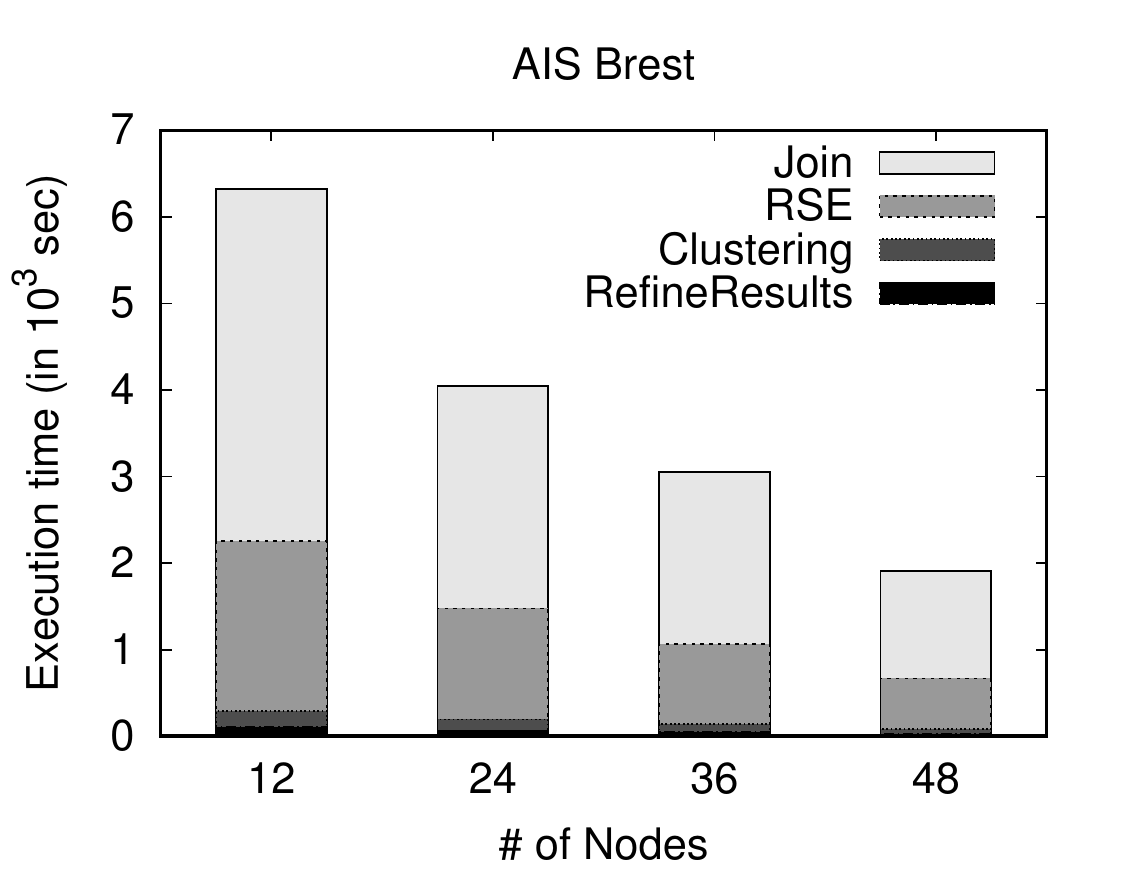}
  \caption{}
\end{subfigure}
~
\begin{subfigure}[t]{0.48\columnwidth}
  \includegraphics[width=\columnwidth]{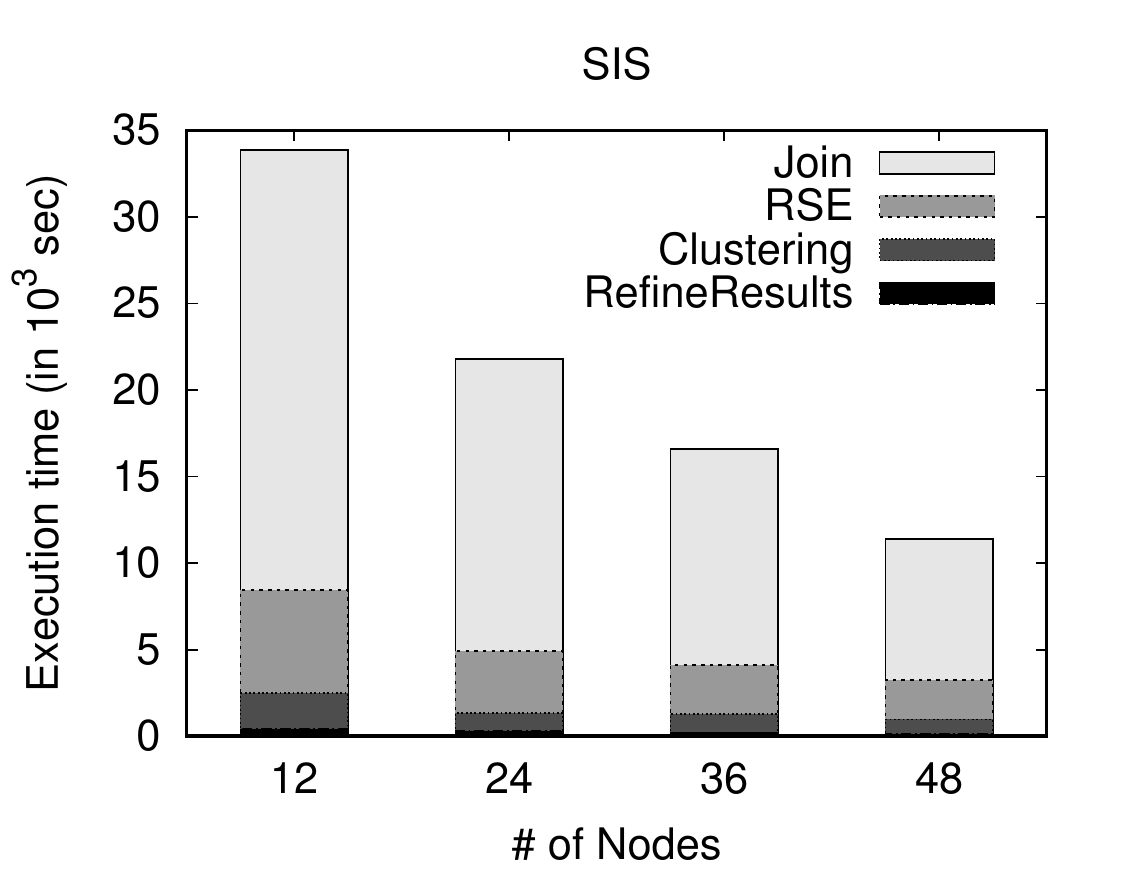}
  \caption{}
\end{subfigure}
 \caption{Scalability analysis varying the size of the (a) AIS Brest and (b) SIS dataset and the number of nodes over the (c)AIS Brest and (d) SIS dataset}
\label{fig_scalab}
\end{figure*}

\subsection{Performance and Scalability} \label{sec_scalab}

\begin{figure*}[thb]
\begin{subfigure}[t]{0.48\columnwidth}
  \includegraphics[width=\columnwidth]{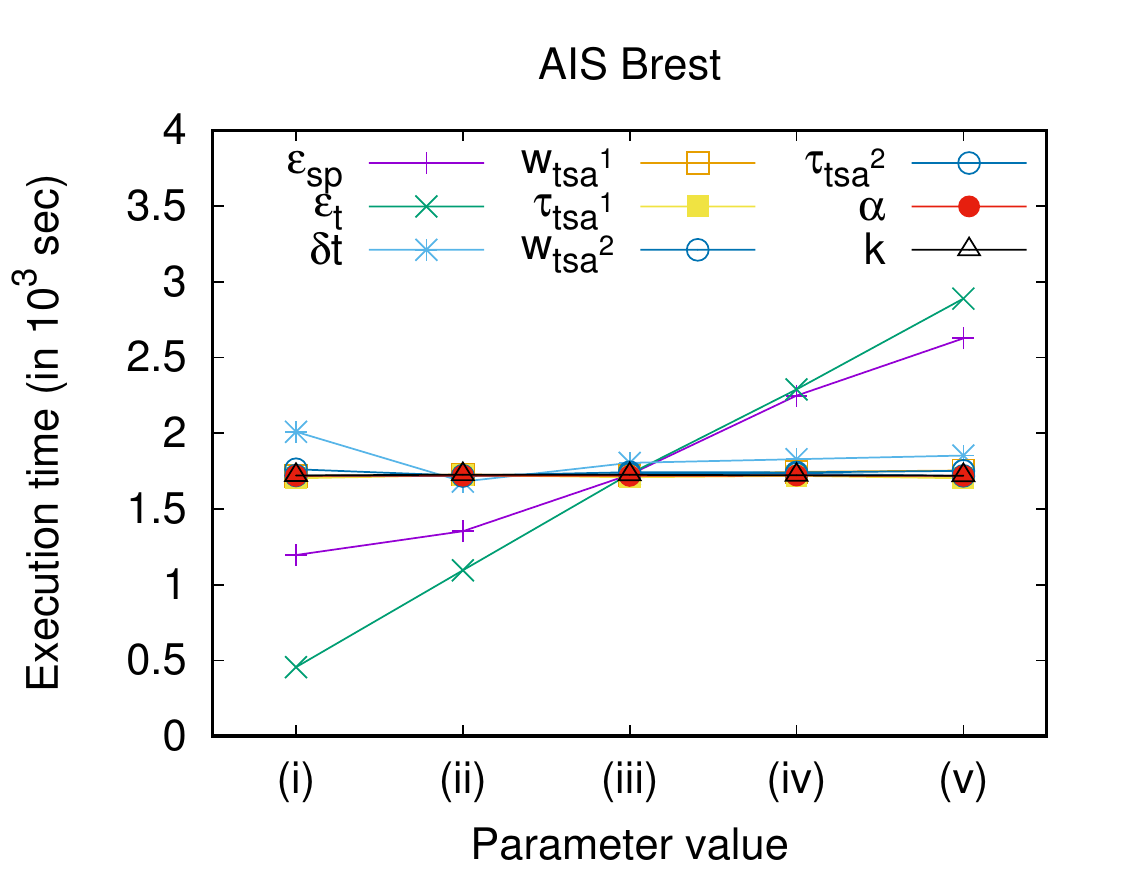}
  \caption{}
\end{subfigure}
~
\begin{subfigure}[t]{0.48\columnwidth}
  \includegraphics[width=\columnwidth]{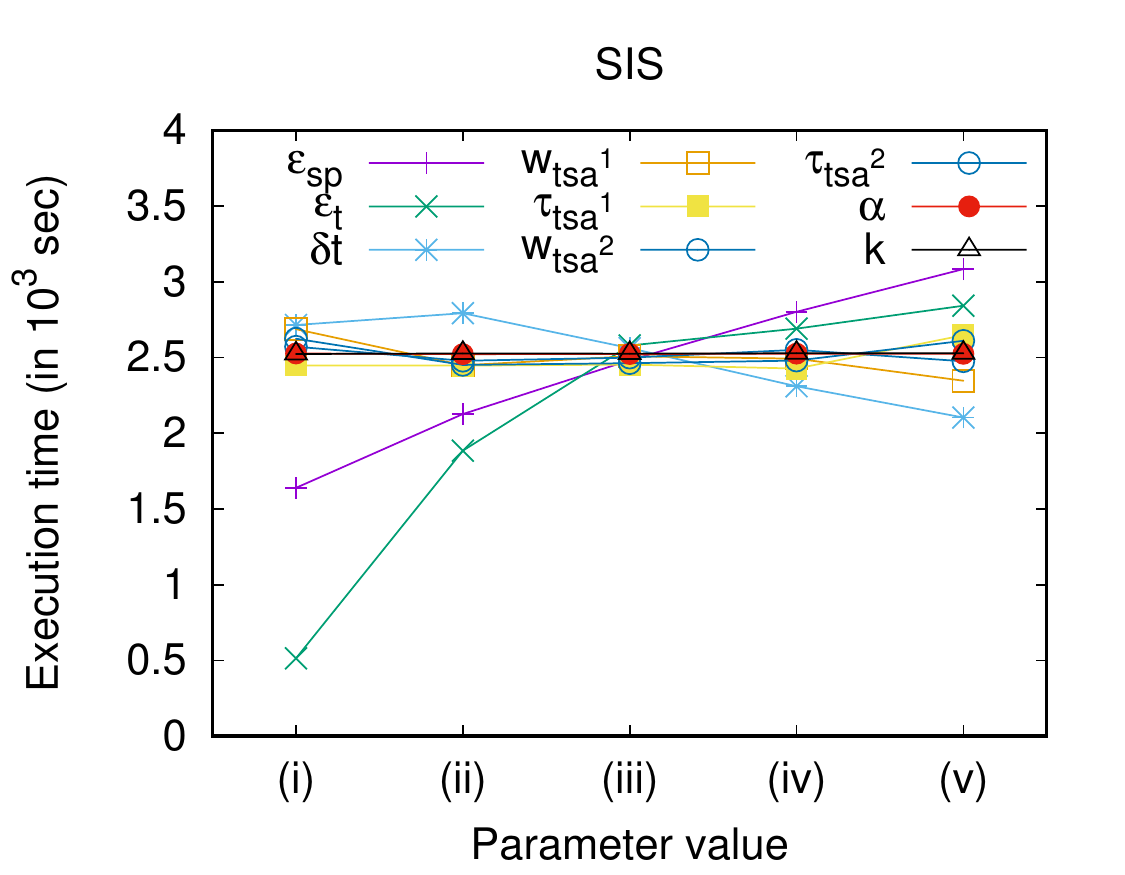}
  \caption{}
\end{subfigure}
~
\begin{subfigure}[t]{0.48\columnwidth}
  \includegraphics[width=\columnwidth]{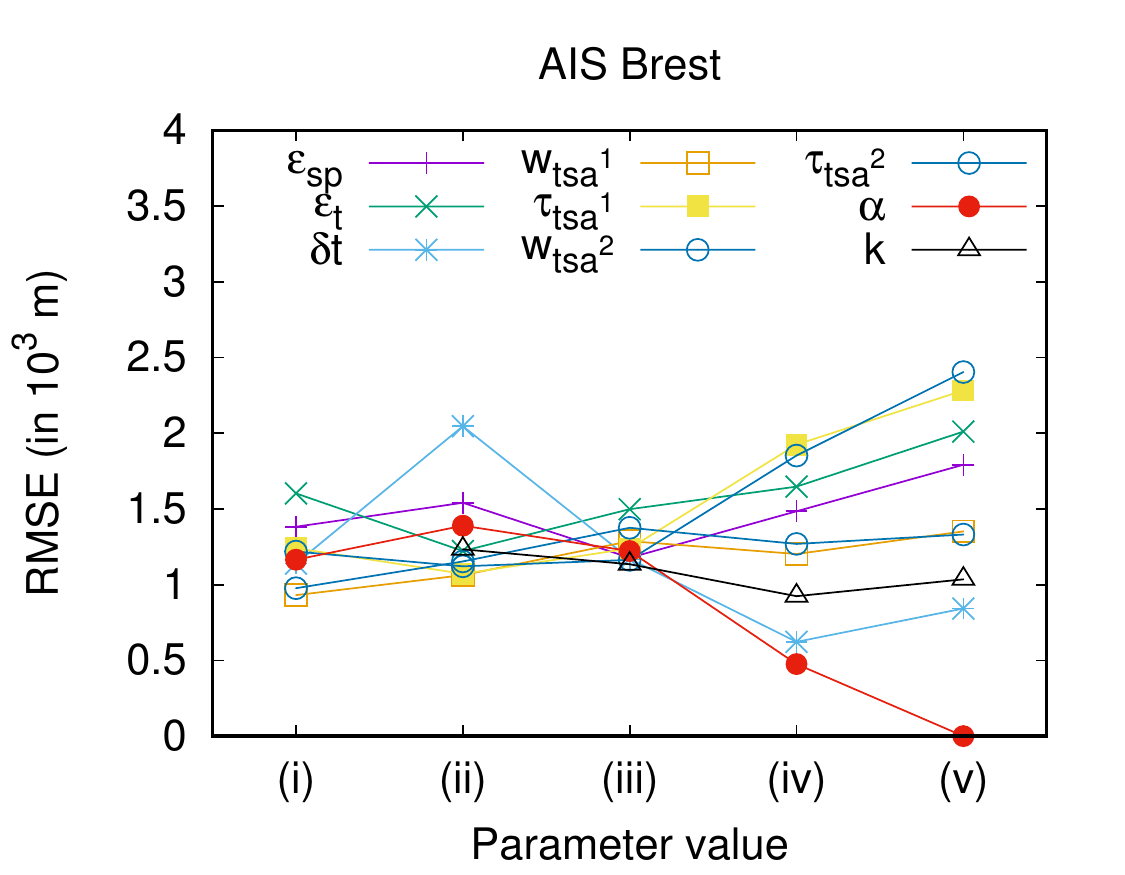}
  \caption{}
\end{subfigure}
~
\begin{subfigure}[t]{0.48\columnwidth}
  \includegraphics[width=\columnwidth]{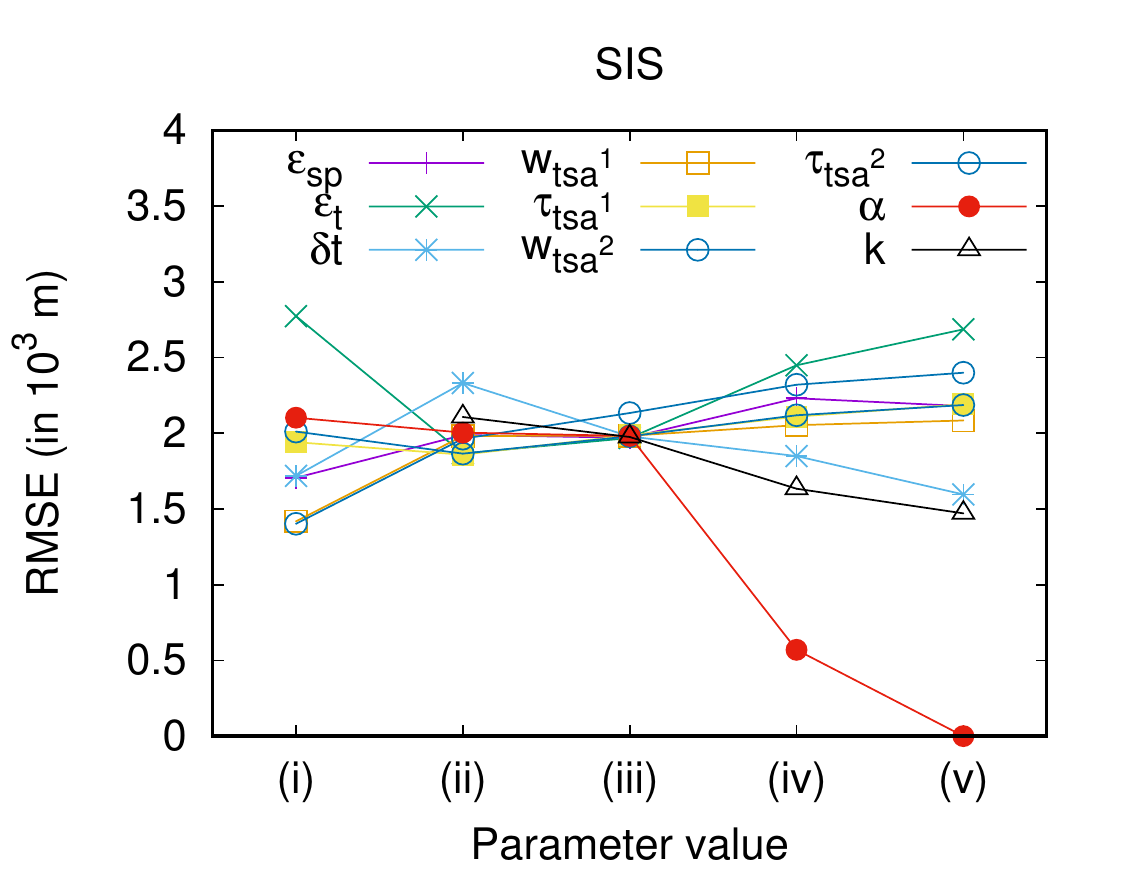}
  \caption{}
\end{subfigure}
\caption{Sensitivity analysis in terms of execution time of (a) the AIS Brest and (b) the SIS dataset and in terms of \emph{RMSE} of (c) the AIS Brest and (d) the SIS dataset}
\label{fig_sens}
\end{figure*}

Initially, we vary the size of our datasets and measure the execution time of our algorithms. We show the impact of the individual steps:  \emph{Join}, \emph{RSE}, \emph{Clustering} and \emph{RefineResult} using stacked bars.  
To study the effect of dataset size, we created 4 portions (20\%, 40\%, 60\%, 80\%) of the original datasets. \emph{RSE} stands for
the \emph{Refine} and \emph{Segmentation} procedure (Figure~\ref{fig_DSC}, Job 1, Reduce phase). 
As illustrated in Figures~\ref{fig_scalab}(a) and (b), as the size of the dataset increases, \emph{DSC} appears to scale linearly. Subsequently, we keep the size of the datasets fixed (at 100\%) and vary the number of nodes. As the number of nodes increases and the dataset size remains the same, it is expected that the execution time will decrease. Indeed, as depicted in Figures~\ref{fig_scalab}(c) and (d), as the number of nodes increases, \emph{DSC} presents linear speedup. This linear behaviour, is somehow anticipated due to the fact that the \emph{DSC} approach is dominated by \emph{DTJ}, in terms of execution time, which presents linear speedup, as shown in~\cite{DBLP_journals/corr/abs-1903-07748}.

Investigating further the performance of the different steps of our proposal, we can observe that, as expected, the execution time of the whole procedure is dominated by the Join step (Figure~\ref{fig_DSC}, Job 1, Map phase), followed by \emph{RSE}. Finally, as anticipated, the \emph{Clustering} and the \emph{RefineResults} step (Figure~\ref{fig_DSC}, Job 2) present very good performance, since the computationally intensive part of the similarity matrix calculation has already been done as part of the previous steps. 

\subsection{Sensitivity Analysis} \label{sec_sensit}

In this section, we perform a sensitivity analysis of all the involved parameters. More specifically, we vary each parameter presented in Table~\ref{tab:params}, while keeping the rest of them in their default value (bold), and we measure their effect in the execution time and the quality of the clustering results, in terms of \emph{RMSE}.
Figures~~\ref{fig_sens}(a) and (b) show that the parameters that appear to have a significant impact on execution time are $\epsilon_t$ and $\epsilon_{sp}$.
This is justified from the fact that these parameters actually affect significantly the complexity of the Join step (Figure~\ref{fig_DSC}, Job 1, Map phase), which is the dominant cost of \emph{DSC}. 
Another parameter that seems to have a perceivable effect on the execution time, is $\delta t$, which in fact ``filters'' the results of \emph{DTJ}, thus fewer data reach the next steps.

Regarding the quality of the clustering results, as illustrated in Figure~~\ref{fig_sens}(c) and (d), all the parameters seem to have an effect over it. In more detail, the larger the values of $\epsilon_t$ and $\epsilon_{sp}$, the larger the \emph{RMSE}. This behaviour is expected since we allow objects that are further away from a representative to participate to the same cluster. In contrast, as $\delta t$ increases, the \emph{RMSE} decreases, which is also anticipated since it sets a lower bound to the longest common subsequence. Furthermore, all the parameters that control the segmentation have the same effect on the \emph{RMSE}, i.e. the smaller (in length) the subtrajectories, the smaller the \emph{RMSE}. This shows that breaking trajectories to subtrajectories has a positive effect on the quality of the clustering and justifies the motivation of our work. Moreover, as $\alpha$ increases the \emph{RMSE} decreases, since for small values of $\alpha$, less similar objects are allowed to participate in a cluster. Finally, the larger the $k$ the smaller the \emph{RMSE}, since it disallows the identification of clusters with small support.

\section{Conclusions} \label{sec_concl}
In this paper, we addressed the problem of \emph{\prob} by building upon a scalable subtrajectory join query operator in order to tackle the problem in an efficient manner. Subsequently, we proposed two alternative trajectory segmentation algorithms. Finally, we proposed a distributed clustering algorithm where the clusters are identified in a parallel manner and get refined as a final step. Our experimental study was performed on a synthetic and two large real datasets of trajectories from the urban and the maritime domain. As for future work, we plan to extend our solution with properties of density-based clustering algorithms. Furthermore, since our algorithm employs a single pass from the data we will investigate the possibility of addressing the same problem in a streaming environment.

\section{Acknowledgements} \label{sec_ack}
This work was partially supported by projects datACRON (grant agreement No 687591), Track\&Know (grant agreement No 780754) and MASTER (Marie Sklowdoska-Curie agreement N. 777695), which have received funding from the EU Horizon 2020 R\&I Programme.

\bibliographystyle{abbrv}

\bibliography{main}
\appendix

\section{Appendix}
\subsection*{Proof of Lemma~\ref{lem_rmse}}

\begin{proof}
\begin{flalign*}
Sim(r', s') = \frac{\sum\limits_{k = 1}^{min(|r'|,|s'|)}(1 - {\frac{d_{s}(r'_k, s'_k)}{\epsilon_{sp}}})}{min(|r'|,|s'|)} \\
Sim(r', s') = \frac{min(|r'|,|s'|)-\sum\limits_{k = 1}^{min(|r'|,|s'|)}{\frac{d_{s}(r'_k, s'_k)}{\epsilon_{sp}}}}{min(|r'|,|s'|)} \\
But,\ \sum\limits_{k = 1}^{min(|r'|,|s'|)}{d_{s}(r'_k, s'_k)} = min(|r'|,|s'|)\cdot\overline{d_{s}(r',s')} \\
So,\ Sim(r', s') = \frac{min(|r'|,|s'|)-\frac{min(|r'|,|s'|)\cdot\overline{d_{s}(r',s')}}{\epsilon_{sp}}}{min(|r'|,|s'|)} \\
Sim(r', s') = \frac{min(|r'|,|s'|)\cdot(1 - \frac{\overline{d_{s}(r',s')}}{\epsilon_{sp}})}{min(|r'|,|s'|)} \\
Sim(r', s') = 1 - \frac{\overline{d(r',s')}}{\epsilon_{sp}} \\
But,\ Sim(r', s') \geq \alpha \\
So,\ \overline{d_{s}(r',s')} \leq \epsilon_{sp}\cdot(1 - \alpha)
\end{flalign*}
\end{proof}

\end{document}